\documentclass[pra,twocolumn,showpacs,preprintnumbers,amsmath,amssymb]{revtex4-1}

\usepackage{graphicx}
\usepackage{dcolumn}
\usepackage{bm}
\usepackage{color}

\usepackage{enumerate}

\def\rr{{\bm r}}

\begin{document}

\preprint{APS/123-QED}


\title{Symmetry Classification of Spinor Bose-Einstein Condensates
}
\author{Yuki Kawaguchi$^1$}
\author{Masahito Ueda$^{1,2}$}
\affiliation{
$^1$Department of Physics, University of Tokyo, 7-3-1 Hongo, Bunkyo-ku, Tokyo 113-0033, Japan\\
$^2$Macroscopic Quantum Control Project, ERATO, JST, Bunkyo-ku, Tokyo 113-8656, Japan
}
\date{\today}

\begin{abstract}
We propose a method for systematically finding ground states
of spinor Bose-Einstein condensates by utilizing symmetry properties of the system.
By this method, we can find not only an inert state, whose symmetry is maximal in the manifold under consideration,
but also a non-inert state, which has lower symmetry and depends on the parameters in the Hamiltonian.
We establish the symmetry-classification method for the
spin-1, 2 and 3 cases at zero magnetic field,
and find a new phase in the last case.
Properties of vortices in the spin-3 system are also discussed.
\end{abstract}

\pacs{03.75.Mn, 05.30.Jp, 03.75.Hh}


\maketitle

\section{introduction}

Classification of ordered states based on symmetries has been employed in many areas of physics, chemistry, and mathematics.
In ordered states, symmetries of the system at high temperatures are spontaneously broken~\cite{LandauLifshitz_SP}.
There exist several phases in quantum condensed systems with internal degrees of freedom,
such as unconventional superconductors, superfluid Helium three, and Bose-Einstein condensates (BECs) with spin degrees of freedom.
The last ones are referred to as spinor BECs.
Symmetries of the system are not completely broken in these phases, and
once we know which symmetry is broken in the ground-state phase,
we can immediately find what types of topological excitations, such as vortices, monopoles, and skyrmions, can be hosted in that phase~\cite{Mermin1979, Michel1980,Mineev1998}.
In this paper, we discuss how to find the ground state of a BEC with internal degrees of freedom,
from a point of view of symmetry classification.

Here, we briefly explain the concept of the symmetry-classification method for the case of spinor BECs.
In the mean-field approximation, we assume that all atoms are Bose-Einstein condensed in a single-particle state, that is the order parameter of the system.
For a spin-$F$ system, the order parameter is a $(2F+1)$-component complex spinor:
$\bm\psi=(\psi_F, \psi_{F-1},\cdots,\psi_{-F})^T$, where $T$ denotes the transpose.
Then, the ground-state order parameter is obtained by minimizing the mean-field energy functional $E[\bm\psi]$ with respect to $\psi_m^*$, i.e., 
\begin{align}
 \frac{\delta E[\bm\psi]}{\delta\psi_m} = 0,\ \ m=F, F-1, \cdots, -F.
\label{eq:stationary}
\end{align}
The above set of equations gives the multi-component Gross-Pitaevskii equation for a stationary state.
In general, to find the ground state, we must solve a set of $2F+1$ nonlinear coupled equations.
Though this procedure works well when $F$ is small~\cite{Ohmi1998,Ho1998,Koashi2000,Ciobanu2000,Ueda2002},
the calculation becomes very involved for large $F$~\cite{Diener2006,Santos2006}.

One reason for the complexity of the problem is that there are an infinite number of solutions to Eq.~\eqref{eq:stationary} associated with symmetry breaking.
In general, a system under consideration has a certain symmetry, which is spontaneously broken in the ordered phase.
For the case of spinor gases, the Hamiltonian is invariant under the global gauge transformation, global spin rotation, and time reversal.
The mean-field energy is also invariant under these transformations on $\bm\psi$.
If we find a solution to Eq.~\eqref{eq:stationary},
the order parameters obtained by applying the gauge transformation, spin rotation, and/or time reversal to the solution
also satisfy Eq.~\eqref{eq:stationary}.
A set of order parameters obtained by such transformations is called an {\it orbit}.
For example, the direction of the spontaneous magnetization in a ferromagnetic state is arbitrary in the absence of an external field.
All ferromagnetic states having different directions of magnetization belong to the same orbit and should be identified as the same class of states.

A set of operations under which the order parameter remains invariant constitutes an {\it isotropy group}.
The isotropy group characterizes the symmetry of the state, and its conjugacy class provides a convenient label to classify the individual states according to their symmetries.
Moreover, such a symmetry consideration gives further clues for finding the ground-state order parameters.
According to Michel~\cite{Michel1971,Michel1980}, the gradient of the energy functional with respect to the order parameter vanishes
in the direction along which the order parameter changes its symmetry.
It follows that if we find a stationary solution 
by restricting the order parameter space so that the order parameter has a certain symmetry,
the obtained state is always stationary in the whole order parameter space.
This theorem greatly simplifies the procedure for finding a stationary state
in comparison with direct solution of Eq.~\eqref{eq:stationary}.
In particular, in some cases, there is only one solution (orbit) which has a certain symmetry.
Such a state, which is called an {\it inert state}, is always stationary and robust against a change in interaction parameters.
Inert states have been obtained from the symmetry consideration in {\it p}- and {\it d}-wave superconductors~\cite{Volovik1985,Ozaki1985}, superfluid helium three~\cite{Bruder1986}, and
spinor BECs~\cite{Makela2007a, Yip2007}.
On the other hand, a {\it non-inert state} depends on the parameters in the interaction energy, and energy minimization must be invoked to find it.

In this paper, we discuss the symmetry-classification method and apply it to spinor BECs with spin-1, 2, and 3 bosons at zero magnetic field.
For the cases of spin-1 and 2 BECs, the obtained results agree with those found in the previous works~\cite{Ohmi1998,Ho1998,Koashi2000,Ciobanu2000,Ueda2002}.
For the case of a spin-3 BEC, however, the systematic method enables us to find a new phase, which exists in a very narrow region in the space of the scattering lengths and 
has eluded the previous works~\cite{Diener2006, Barnett2007}.
In the spin-3 system, there are many phases that have discrete symmetries, 
leading to various kinds of vortices as in the case of a half-quantum vortex in the spin-1 polar phase~\cite{Zhou2001} and a 1/3-vortex in the spin-2 cyclic phase~\cite{Makela2003,Semenoff2007}.
We discuss properties of vortices in spin-3 BECs and show that  the quantization unit of the mass circulation depends on the interaction parameters in some phases.
The study on a spin-3 BEC has been motivated by the experimental realization of BEC of $^{52}$Cr atoms~\cite{Griesmaier2005,Beaufils2008}.
Examples include the phase diagrams in the presence or absence of an external field~\cite{Diener2006,Santos2006}, 
those under a light-induced quadratic Zeeman energy~\cite{Santos2007}
or conserved magnetization~\cite{Makela2007b},
and phase separation under an external magnetic field~\cite{He2009}.
Possible vortices in each phase are investigated in Ref.~\cite{Barnett2007}.
Recently, spinor properties of spin-3 $^{52}$Cr BEC have been observed~\cite{Pasquiou2011,Pasquiou2011b}.

This paper is organized as follows.
In Sec.~\ref{sec:method}, we describe a method for finding a stationary point of an arbitrary function on a smooth manifold,
and establish mathematical notations used in the present paper.
In Sec.~\ref{sec:procedure}, we describe a general procedure of the symmetry-classification method in spinor BECs.
In Sec.~\ref{sec:examples}, we carry out this procedure for spin-1, 2, and 3 BECs.
In particular, for the case of a spin-3 BEC, we point out a new phase which has eluded Refs.~\cite{Diener2006, Barnett2007}.
In Sec.~\ref{sec:examples}, we discuss properties of vortices in spin-3 BECs.
In Sec.~\ref{sec:conclusions}, we make concluding remarks.
In the appendix, we explore stationary states with discrete symmetries in spin-3 BECs.

\section{Symmetry-Classification Method}
\label{sec:method}

Our symmetry-classification method is based on the following Michel's theorems~\cite{Michel1971,Michel1980,Vollhardt1990}.

\subsection{Michel's Theorem}

We consider a real smooth function $f$ on a smooth manifold $\mathcal{M}$.
Let $G$ be a group of operations which automorphically map $\mathcal{M}$ to itself and do not change the value of $f$:
\begin{align}
G=\{g\in {\rm Aut}\mathcal{M}|f(g x) = f(x)\ \textrm{for}\ ^\forall x\in\mathcal{M}\},
\end{align}
where ${\rm Aut}\mathcal{M}$ denotes the group of automorphisms on $\mathcal{M}$.
For spinor BECs,
$f$ is the mean-field energy, $\mathcal{M}$ is the order-parameter manifold,
and $G$ is a group of gauge transformations, spin rotations, and time reversal (see Sec.~\ref{sec:mean-field-energy}).

An {\it orbit} $\mathcal{M}_O(x)$ of $x\in \mathcal{M}$ is defined as the trajectory of a point $x$ on the manifold under $G$:
\begin{align}
\mathcal{M}_O(x) = \{g x \in \mathcal{M}|g\in G\} \subset \mathcal{M}.
\end{align}
By assumption, $f$ takes on the same value on all the points in $\mathcal{M}_O(x)$.
If $x$ is a stationary point in $f$, $\mathcal{M}_O(x)$ is also stationary.
What we need to find is not a stationary point $x$, but a stationary orbit $\mathcal{M}_O(x)$.
An {\it isotropy group} $H_x$ is a set of operations that do not change $x$:
\begin{align}
H_x = \{g\in G | gx=x\} \subset G.
\end{align}
It is clear that $H_x$ is a subgroup of $G$.
It can also be shown that the isotropy groups of points
on the same orbit are conjugate to each other:
\begin{align}
 H_{gx}&=\{g'\in G|g'gx=gx\}\nonumber\\
&=\{g'\in G|g^{-1}g'gx=x\}\nonumber\\
&=\{gg''g^{-1}\in G|g''\in H_x\}\nonumber\\
&=gH_xg^{-1}.
\end{align}
Here, $H$ and $H'$, which are subgroups of $G$, are conjugate to each other if and only if there exists $g\in G$ such that $H'=gHg^{-1}$.
If two points on different orbits share the same isotropy group,
the orbits of such two points are considered to be of the same type,
and we classify the types of orbits according to the conjugacy classes of subgroups of $G$.
In other words, for each conjugacy class of a subgroup of $G$, we obtain a set of orbits.
Such a union of orbits is called a {\it stratum} $\mathcal{M}_S(x)$;
$x$ and $x'\in \mathcal{M}$ belong to the same stratum, if and only if their isotropy groups are conjugate to each other.
Clearly, $\mathcal{M}_O(x)\subseteq \mathcal{M}_S(x)\subset \mathcal{M}$.

Embedding the manifold $\mathcal{M}$ in an $n$-dimensional Euclidean space $\mathbb{R}^n$, where $n\ge{\rm dim}\mathcal{M}$,
the gradient of $f$ at $x=(x_1,x_2,\cdots,x_n)$ is defined as
\begin{align}
\nabla_\mathcal{M}f(x)=\left(\frac{\partial f}{\partial x_1},\frac{\partial f}{\partial x_2},\cdots,\frac{\partial f}{\partial x_n}\right).
\end{align}
Here, $\nabla_\mathcal{M}f(x)$ describes the direction of the steepest-ascent vector which is
tangent to the manifold $\mathcal{M}$.
Michel has proved that $\nabla_\mathcal{M}f(x)$ is tangent to the stratum $\mathcal{M}_S(x)$, i.e.,
the gradient of $f$ vanishes in the direction along which the symmetry of the state changes~\cite{Michel1971,Michel1980}.
Moreover, since $f$ is a $G$-invariant function, $\nabla_\mathcal{M}f(x)$ is zero in the direction of $\mathcal{M}_O(x)$.
Hence, we obtain the following theorems:

\vspace{2mm}
\noindent{\bf Theorem 1 (inert state).}
If an orbit is isolated in the stratum, the orbit is stationary.

\vspace{2mm}
\noindent{\bf Theorem 2 (non-inert state).}
If an orbit is not isolated in the stratum, 
we define a submanifold $\mathcal{M}_H\subset \mathcal{M}$ such that
\begin{align}
 \mathcal{M}_H=\{x\in \mathcal{M}| hx=x\ \textrm{for}\ ^\forall h\in H\},
\end{align}
where $H$ is a subgroup of $G$ that characterizes the stratum under consideration.
Let $f_H$ be a real function which is the same as $f$ but whose domain is restricted on $\mathcal{M}_H$.
Then, the stationary point of $f_H$ on $\mathcal{M}_H$ is always a stationary point of $f$ on $\mathcal{M}$.

\vspace{2mm}
\noindent
It follows from Theorem 1 that all $G$-invariant functions $f$ on a manifold $\mathcal{M}$ have a common stationary orbit.
The corresponding state is called an inert state.
Theorem 2 is instrumental in finding non-inert states.

\subsection{Procedure}
\label{sec:Michel_Procedure}
Following the above two theorems, the procedure to find a minimum of a $G$-invariant function $f$ on a manifold $\mathcal{M}$ is summarized as follows:
\begin{enumerate}[{\bf 1.}]
\item Classify all subgroups of $G$ according to conjugacy classes.
\item Let $H$ be an element of a conjugacy class.
Find such $x\in \mathcal{M}$ that it is invariant under $H$.
\item \label{item:inert} (inert state) If $x$ is uniquely determined, then $x$ is a stationary point of $f$, and the corresponding orbit $\mathcal{M}_O(x)$ is a stationary orbit.
\item \label{item:noninert} (non-inert state) If $x$ is not uniquely determined,
      then calculate the minimum of $f$ in the submanifold $\mathcal{M}_H\equiv \{ x\in\mathcal{M}|hx=x\ \textrm{for}\ ^\forall h\in H\}$.
      A stationary point $x\in\mathcal{M}_H$ is also a stationary point of $f$ in the whole space of $\mathcal{M}$.
\item Finally, compare the values of $f$ for the obtained stationary states and find the lowest one.
\end{enumerate}

We emphasize that the above procedure works well for the case in which stationary states have a certain symmetry.
If $H=1$, i.e., if no symmetry remains, the above procedure amounts to solving Eq.~\eqref{eq:stationary} directly.
We therefore do not consider the case of $H=1$.
In the absence of an external field,
all ground states of spinor BECs with spin $F=1,2$ and 3, superfluid $^3$He, and {\it p}- and {\it d}-wave superconductors
have remaining symmetries.

\section{General Procedures for Symmetry Classification of Spinor Condensates}
\label{sec:procedure}
In this section, we describe general procedures for applying the symmetry-classification method described in the preceding section to the case of spinor BECs.

\subsection{Mean-field energy}
\label{sec:mean-field-energy}
The mean-field energy of a uniform system of spin-$F$ atoms with mass $M$ at zero magnetic field is given by
\begin{align}
E[\bm\psi] &= -\frac{\hbar^2}{2M}\int d\rr \sum_m\psi_m^* \nabla^2\psi_m\nonumber\\
&+ \frac{1}{2}\iint d\rr d\rr' \sum_{mnm'n'} V^{mm'}_{nn'}(\rr-\rr') \nonumber\\
&\hspace{20mm}\psi_m^*(\rr)\psi_{m'}^*(\rr')\psi_{n'}(\rr')\psi_{n}(\rr).
\end{align}
We consider only a short-range interaction for $V^{mm'}_{nn'}(\rr-\rr')$, and ignore the magnetic dipole-dipole interaction.
Then, the interaction potential conserves the total spin of two colliding atoms
and can be approximated with the delta function as
\begin{align}
&V^{mm'}_{nn'}(\rr-\rr')=\delta(\rr-\rr') C^{mm'}_{nn'},\\
&C^{mm'}_{nn'}=\sum_{S=0,{\rm even}}^{2F} \frac{4\pi\hbar^2}{M}a_S \langle Fm,Fm'|\mathcal{P}_S|Fn,Fn'\rangle,
\end{align}
where $a_S$ is the {\it s}-wave scattering length of the total spin $S$ channel, and
$\mathcal{P}_S=\sum_{M_S}|S,M_S\rangle \langle S,M_S|$ projects a pair of atoms onto the total spin $S$ state.
From the fact that the inter-atomic interaction is elastic and conserves the total number and total spin of particles,
the Hamiltonian is invariant under the global $U(1)$ gauge transformation, the $SO(3)$ rotation in spin space,
and time reversal $\Theta$
\footnote{
Although the Hamiltonian is also invariant under spin inversion, it reduces to time reversal since
$\sum_{mn}(\mathcal{T}\zeta)_m^*(\bm F)_{mn}(\mathcal{T}\zeta)_n = -\sum_{mn} \zeta_m^*(\bm F)_{mn}\zeta_n$, where $\mathcal{T}$ is the time-reversal operator defined in Eq.~\eqref{eq:time_reversal_operator}}.
Hence,
\begin{align}
G=U(1)_\phi \times SO(3)_F \times \Theta
\label{eq:G}
\end{align}
is the full symmetry of spinor gases.
Here, the subscripts $\phi$ and $F$ denote the gauge and spin symmetry, respectively.
If the scattering lengths satisfy special relations, $G$ can be enlarged.
For example, if all $a_S$'s are equal, $G=SU(2F+1)\times \Theta$.
In this paper, however, we do not consider such exceptions.

In the absence of a trapping potential or a long-range interaction, the ground state is uniform with fixed density $n$.
Introducing a normalized spinor $\bm\zeta = (\zeta_F, \zeta_{F-1},\cdots,\zeta_{-F})^T$ as $\psi_m = \sqrt{n}\zeta_m$,
the ground state is obtained by minimizing 
\begin{align}
 \mathcal{E}[\bm\zeta] \equiv \frac{2E[\psi]}{n^2\Omega}= \sum_{mnm'n'}C^{mm'}_{nn'} \zeta_m^*\zeta_{m'}^*\zeta_{n'}\zeta_{n},
\label{eq:mathcalE}
\end{align}
subject to the normalization condition $\bm\zeta^\dagger\bm\zeta = 1$,
where $\Omega$ is the volume of the system.
Using relations
$\sum_S\mathcal{P}_S=1$ and 
$\sum_S [S(S+1)-2F(F+1)]\mathcal{P}_S=2\bm F\cdot \bm F$~\cite{Ho1998},
Eq.~\eqref{eq:mathcalE} for $F=1, 2$ and 3 can be rewritten as
\begin{align}
\mathcal{E}^{(1)}[\bm\zeta] &= c_0 + c_1|\bm f|^2,\\
\mathcal{E}^{(2)}[\bm\zeta] &= c_0 + c_1|\bm f|^2 + c_2|A_{00}|^2,\\
\mathcal{E}^{(3)}[\bm\zeta] &= c_0 + c_1|\bm f|^2 + c_2|A_{00}|^2 + c_3 \sum_M |A_{2M}|^2,\label{eq:E3}
\end{align}
respectively,
where 
\begin{align}
\bm f &= \sum_{mn}\zeta_m^* \bm F_{mn}\zeta_n,\\
A_{00} &= \sum_{mn}\langle 0,0|Fm,Fn\rangle \zeta_m\zeta_{n},\\
A_{2M} &= \sum_{mn}\langle 2,M|Fm,Fn\rangle \zeta_m\zeta_{n}, 
\end{align}
are the magnetization per particle, the spin-singlet pair amplitude, and the spin-quintet pair amplitude, respectively, 
with $\bm F=(F_x,F_y,F_z)$ being the vector of spin-$F$ matrices and $\langle S,M|Fm,Fn\rangle$ the Clebsch-Gordan coefficient.
The coupling constants $c_0, c_1, c_2$ and $c_3$ are given in terms of the scattering lengths as~\cite{Ho1998,Koashi2000,Ciobanu2000,Ueda2002,Diener2006,Santos2006}
\begin{align}
 F=1:&\ \bar{c}_0=2a_2+a_0,\nonumber\\
     &\ \bar{c}_1=a_2-a_0,\\
 F=2:&\ \bar{c}_0=3a_4+4a_2,\nonumber\\
     &\ \bar{c}_1=a_4-a_2,\nonumber\\
     &\ \bar{c}_2=3a_4-10a_2+7a_0,\\
 F=3:&\ \bar{c}_0=2a_6+9a_4,\nonumber\\
     &\ \bar{c}_1=a_6-a_4,\nonumber\\
     &\ \bar{c}_2=10a_6-21a_4+11a_0,\nonumber\\
     &\ \bar{c}_3=7a_6-18a_4+11a_2,
\end{align}
where $\bar{c}_i\equiv (4F-1)Mc_i/(4\pi\hbar^2)$.

For $F=3$, using the identity
$\sum_S[S(S+1)-2F(F+1)]^2\mathcal{P}_S=(2\bm F\cdot \bm F)^2$,
the last term on the right-hand side of Eq.~\eqref{eq:E3} can be rewritten as
\begin{align}
\sum_M|A_{2M}|^2 =& -\frac{1}{7}-\frac{1}{18}|\bm f|^2 -\frac{5}{3}|A_{00}|^2 \nonumber\\
 &+ \frac{1}{126}\left[\sum_{\mu\nu=x,y,z}\sum_{mn}\zeta_m^*(F_{\mu}F_{\nu})_{mn}\zeta_n\right]^2.
\label{eq:A2M}
\end{align}
By decomposing $F_\mu F_\nu$ in the last term into the symmetric and antisymmetric parts:
\begin{align}
F_{\mu}F_{\nu} 
&= \frac{F_{\mu}F_{\nu}+F_{\nu}F_{\mu}}{2} +\frac{i}{2}\sum_\lambda\epsilon_{\mu\nu\lambda}F_\lambda,
\end{align}
Eq.~\eqref{eq:A2M} reduces to
\begin{align}
\sum_M|A_{2M}|^2 =& -\frac{1}{7}-\frac{5}{84}|\bm f|^2 -\frac{5}{3}|A_{00}|^2 
 + \frac{1}{126}{\rm Tr}\,\mathcal{N}^2,
\end{align}
where $\mathcal{N}$ is the spin nematic tensor defined by~\cite{Diener2006}
\begin{align}
 \mathcal{N}_{\mu\nu} = \sum_{mn}\zeta_m^* \left(\frac{F_\mu F_\nu+F_\nu F_\mu}{2}\right)_{mn}\zeta_n,\ \ \mu,\nu=x,y,z.
\label{eq:def_mathcalN}
\end{align}
By definition, $\mathcal{N}$ is a real symmetric tensor whose trace is given by ${\rm Tr}\,\mathcal{N}=F(F+1)$.
Then, the mean-field energy for $F=3$ [Eq.~\eqref{eq:E3}] is rewritten as
\begin{align}
\mathcal{E}^{(3)}[\bm\zeta] &= \tilde{c}_0 + \tilde{c}_1|\bm f|^2 + \tilde{c}_2|A_{00}|^2 + \tilde{c}_3 {\rm Tr}\,\mathcal{N}^2,\label{eq:E3'}
\end{align}
where $\tilde{c}_i$'s are related to $c_i$'s as
\begin{align}
 \tilde{c}_0 &= c_0 - \frac{1}{7} c_3, \ \ \
 \tilde{c}_1 = c_1 - \frac{5}{84} c_3, \nonumber\\ 
 \tilde{c}_2 &= c_2 - \frac{5}{3} c_3,\ \ \ 
 \tilde{c}_3 = \frac{1}{126} c_3. \label{eq:cc3}
\end{align}

\subsection{Procedure for the case of spinor BECs}
Following the procedure in Sec.~\ref{sec:Michel_Procedure},
we first classify all subgroups of $G$ given by Eq.~\eqref{eq:G} according to conjugacy classes.
The representation of each component of $G$ in the spin-$F$ manifold is given by
\begin{align}
 U(1)_\phi&=\{e^{i\phi}{\bf 1}|\phi\in\mathbb{R} \},\\
 SO(3)_F  &=\{e^{-i F_z\alpha}e^{-iF_y\beta}e^{-iF_z\gamma}|\alpha,\beta,\gamma\in\mathbb{R} \},\\
 \Theta   &=\{{\bf 1}, \mathcal{T}\},
\end{align}
where ${\bf 1}$ is the $(2F+1)\times (2F+1)$ identity matrix, 
$\alpha, \beta$ and $\gamma$ are Euler angles,
and the time-reversal operator $\mathcal{T}$ acts on $\bm\zeta$ as
\begin{align}
 (\mathcal{T} \bm\zeta)_{m} = (-1)^m\zeta_{-m}^*.
\label{eq:time_reversal_operator}
\end{align}

Since the eigenvalue of an arbitrary element $g$ of $SO(3)$ is given in the form of $e^{i\lambda}$ with $\lambda\in \mathbb{R}$, 
the corresponding eigenstate is invariant under $e^{-i\lambda}g\in U(1)_\phi\times SO(3)_F$,
that is, it is invariant under a spin-gauge coupled operation.
If there are two eigenstates that have different eigenvalues, these two states differ in the spin-gauge symmetry.
Hence, the procedure 1 and 2 in Sec.~\ref{sec:Michel_Procedure} are rephrased as follows:
\begin{enumerate}[{\bf 1'.}]
\item List all subgroups of $SO(3)_F$.
\item \label{item:eigenstate}Let $H'$ be a subgroup of $SO(3)_F$.
      Find simultaneous eigenstates $\bm\zeta$ of all elements of each $H'$.
\end{enumerate}

For procedure {\bf 1'},
it is known that $SO(2)$ is the only continuous subgroup of $SO(3)$,
and that the discrete subgroups of $SO(3)$ are given as follows~\cite{LandauLifshitz_QM}:
\begin{itemize}
\item[$C_n$:] The cyclic group of rotations about a symmetry axis through angle $2\pi k/n$ with $k=1,2,\cdots, n-1$. The group is isomorphic to $\mathbb{Z}_n$.
\item[$D_n$:] The dihedral group generated by the elements of $C_{n}$ and an additional rotation through $\pi$ about an orthogonal axis. 
\item[$T$:] The point group of the tetrahedron composed of 4 three-fold axes and 3 two-fold axes.
\item[$O$:] The point group of the octahedron composed of 3 four-fold axes, 4 three-fold axes, and 6 two-fold axes.
\item[$Y$:] The point group of the icosahedron composed of 6 five-fold axes, 10 three-fold axes, and 15 two-fold axes.
\end{itemize}
Since all rotations through a fixed angle about different axes are conjugate to each other,
we choose a representative element $H'$ in a conjugacy class of each subgroup of $SO(3)_F$ so that the highest symmetry axis is parallel to the $z$ axis.
The generators of such representative elements of the subgroups of $SO(3)_F$ are summarized in Table~\ref{table:generators},
where $C_{n,\omega_x x +\omega_y y + \omega_z z}$ denotes a rotation about the direction $\bm\omega=(\omega_x,\omega_y,\omega_z)$ through $2\pi/n$:
\begin{align}
C_{n,\omega_x x +\omega_y y + \omega_z z} = \exp\left[-i \frac{\omega_x F_x + \omega_y F_y + \omega_z F_z}{|\bm \omega|}\frac{2\pi}{n}\right].
\label{eq:Cnomega}
\end{align}
For example, the matrix elements of $C_{nz}$ and $C_{2x}$ in a spin-$F$ system are given by
\begin{align}
 [C_{nz}]_{mm'} &= e^{-i2\pi m/n}\delta_{mm'},\\
 [C_{2x}]_{mm'} &= (-1)^F\delta_{m,-m'}.
\end{align}

\begin{table}
\begin{tabular}{l|l}\hline
subgroup\hspace{2mm} & generators\\ \hline\hline
$SO(2)$ & $F_z$\\
$C_n$ & $C_{nz}$\\
$D_n$ & $C_{nz}$, $C_{2x}$\\
$T$   & $C_{3z}$, $C_{2,\sqrt{2}x+z}$\\
$O$   & $C_{4z}$, $C_{2,x+z}$ \\
$Y$   & $C_{5z}$, $C_{2,2x+(1+\sqrt{5})z}$ \\ \hline
\end{tabular}
\caption{Generators of the discrete subgroups of $SO(3)$,
where $C_{n,\omega_x x +\omega_y y + \omega_z z}$ denotes a rotation about the direction $\bm\omega=(\omega_x,\omega_y,\omega_z)$ through $2\pi/n$ as given by Eq.~\eqref{eq:Cnomega}.
For example, generator $C_{2,2x+(1+\sqrt{5})z}$ in the icosahedral group denotes a $\pi$-rotation about $\bm \omega=(2,0,1+\sqrt{5})$.
}
\label{table:generators}
\end{table}

Next, we calculate simultaneous eigenstates of all generators of each subgroup $H'$.
First, we consider the case of $H'=SO(2)$. The eigenstate of the generator $F_z$ is uniquely determined as $[\bm\zeta^{(m_0)}]_m\equiv\delta_{m m_0}$.
Here, we neglect the overall phase factor, since $e^{i\phi}\bm\zeta^{(m_0)}\in \mathcal{M}_O(\bm\zeta^{(m_0)})$ for $^\forall\phi\in\mathbb{R}$.
We also disregard negative $m_0$, since 
the order parameter $\bm\zeta^{(-m_0)}$ is obtained by applying the time-reversal or the spin-rotation operator to $\bm\zeta^{(m_0)}$: $\bm\zeta^{(-m_0)}=e^{i\pi F}C_{2x}\bm\zeta^{(m_0)}=e^{i\pi m_0}\mathcal{T}\bm\zeta^{(m_0)}\in \mathcal{M}_O(\bm\zeta^{(m_0)})$.
For $m_0\neq0$, the isotropy group $H^{(m_0)}\subset G$ of $\bm\zeta^{(m_0)}$ is given by
\begin{align}
H^{(m_0)} = \{e^{im_0\alpha}e^{-iF_z\alpha}, e^{i[2m_0\gamma+(F+m_0)\pi]}U^\gamma_2 \mathcal{T}\},
\label{eq:H^m_0}
\end{align}
where 
\begin{align}
 U^\gamma_2\equiv C_{2,(\cos\gamma)x+(\sin\gamma)y} = e^{-iF_z\gamma}C_{2x}e^{iF_z\gamma}
\end{align}
is a $\pi$ rotation about an axis in the $x$--$y$ plane,
and $\alpha$ and $\gamma$ are arbitrary real numbers. 
Here, we have used the following relations:
\begin{align}
  e^{im_0\alpha}e^{-iF_z\alpha}\bm\zeta^{(m_0)} &= \bm\zeta^{(m_0)},\\
 (U_2^\gamma)_{mm'}\zeta^{(m_0)}_{m'} 
&= (e^{-iF_z\gamma}C_{2x}e^{iF_z\gamma})_{mm'}\zeta^{(m_0)}_{m'}\nonumber\\
&=e^{i\pi F} e^{2im_0\gamma}\zeta^{(m_0)}_{-m}\nonumber\\
&= e^{i[2m_0\gamma+(F+m_0)\pi]} (\mathcal{T}\bm\zeta^{(m_0)})_m.
\end{align}
Since $\bm\zeta^{(m_0)}$ and its time-reversal $\mathcal{T}\bm\zeta^{(m_0)}$ are transformed to each other through a spin rotation,
the time-reversal symmetry is broken in these states.
On the other hand, the $m_0=0$ state has the time-reversal symmetry, which is decoupled from spin rotations in the isotropy group:
\begin{align}
H^{(0)} = \{e^{-iF_z\alpha}, e^{i F\pi}U^\gamma_2 \}\times \Theta.
\label{eq:H^m_00}
\end{align}
Note that the $m_0=0$ state also has the $\mathbb{Z}_2$ symmetry~\cite{Zhou2001}:
the order parameter is invariant under the $\pi$ rotation about an axis perpendicular to the $z$ axis,
and the isotropy group can be written as $H^{(0)}=D_\infty \times \Theta$,
where $D_\infty$ denotes the dihedral group of order $n=\infty$.

For the case of discrete subgroups, we shall calculate simultaneous eigenstates and the corresponding isotropy groups for each subgroup $H'$ in Sec.~\ref{sec:examples}.
Most of them are not determined uniquely, and we minimize the energy $\mathcal{E}^{(F)}[\bm\zeta]$ in the submanifold spanned by the simultaneous eigenstate for each $H'$.
We note that the obtained state for a given $H'$ might have higher symmetry than $H'$.
For example, since $C_n\subset D_n$, the stationary point in the submanifold with symmetry $C_n$ may have the $D_n$ symmetry.
To identify the symmetry of the obtained state, it is convenient to use the Majorana representation~\cite{Majorana1932}, which we explain in the next subsection.

\subsection{Majorana representation}

Majorana invented a geometrical representation of a general spin-$F$ state~\cite{Majorana1932}.
A state of the spin-$F$ system can be specified by providing a symmetric configuration of $2F$ spin-1/2 systems, expect for an overall phase.
Since the state of a spin-1/2 system can be described by the unit-sphere Bloch vector,
the state of a spin-$F$ system can be described by $2F$ vertices on the unit sphere.
This representation helps us identify the symmetry structure of a spin-$F$ condensate~\cite{Barnett2006}.

We consider a polynomial of degree $2F$ for a given order parameter $\bm\zeta$:
\begin{align}
P^{(F)}_{\bm \zeta}(z) = \sum_{\alpha=0}^{2F} \sqrt{\begin{pmatrix} 2F \\ \alpha \end{pmatrix}} \zeta_{\alpha-F}^* z^{\alpha}.
\end{align}
Then, the $2F$ complex roots of $P^{(F)}_{\bm \zeta}(z)=0$ give $2F$ vertices on the unit sphere 
through the stereographic mapping $z=\tan(\theta/2)e^{i\phi}$.
For the cases of spin $F=1, 2$ and 3, the polynomials $P^{(F)}_{\bm\zeta}(z)$ are given, respectively, by
\begin{align}
 P^{(1)}_{\bm\zeta}(z) =& \zeta^*_1 z^2 + \sqrt{2} \zeta^*_0 z + \zeta^*_{-1},\\
 P^{(2)}_{\bm\zeta}(z) =& \zeta^*_2 z^4 + 2 \zeta^*_1 z^3 + \sqrt{6}\zeta^*_0 z^2 + 2\zeta^*_{-1} z + \zeta^*_{-2},\\
 P^{(3)}_{\bm\zeta}(z) =& \zeta^*_3 z^6 + \sqrt{6} \zeta^*_2 z^5 + \sqrt{15}\zeta^*_1 z^4 + \sqrt{20} \zeta^*_0 z^3\nonumber\\
 &+ \sqrt{15} \zeta^*_{-1} z^2  + \sqrt{6} \zeta^*_{-2} z +\zeta^*_{-3}.
\end{align}

Using the Majorana representation, we can immediately find some inert states~\cite{Yip2007}.
For example, a spin-2 BEC is described with 4 vertices.
When these 4 vertices form a tetrahedron, the corresponding state has the tetrahedral symmetry.
On the other hand, since an octahedron and an icosahedron have 6 and 12 vertices, respectively,
no spin-2 state has octahedral or icosahedral symmetry.
The octahedral symmetry appears in systems with $F\ge 3$, and the icosahedron symmetry in systems with $F\ge 6$.

For the time-reversed state $\mathcal{T}\bm\zeta$, the polynomial is given by
\begin{align} 
P^{(F)}_{\mathcal{T}\bm \zeta}(z) =& \sum_{\alpha=0}^{2F} \sqrt{\begin{pmatrix} 2F \\ \alpha \end{pmatrix}} (-1)^{F-\alpha}\zeta_{F-\alpha} z^{\alpha}\\
 =& (-1)^Fz^{2F}\sum_{\alpha=0}^{2F} \sqrt{\begin{pmatrix} 2F \\ \alpha \end{pmatrix}} \zeta_{\alpha-F} \left(-\frac{1}{z}\right)^{\alpha}.
\end{align}
If $z=z_0$ is a root of $P^{(F)}_{\bm\zeta}(z)=0$, then $z=-1/z_0^*$ is a root of $P^{(F)}_{\mathcal{T}\bm\zeta}(z)=0$,
which corresponds to the antipole of $z_0$ on the unit sphere.
Hence, the time-reversed state is described with antipoles of vertices of the original state.

\section{Specific Examples in Spinor Bose-Einstein Condensates}
\label{sec:examples}
In this section, we apply the procedure discussed in the previous sections to spin $F=1, 2$, and 3 systems.

\subsection{Spin-1}
\label{sec:spin1}

\subsubsection{Mean-field energy}
A spin-1 spinor BEC is described with a three-component spinor $\bm\zeta=(\zeta_1,\zeta_0,\zeta_{-1})^T$.
The scaled mean-field energy for a given order parameter $\bm\zeta$ is given by
\begin{align}
\mathcal{E}^{(1)}[\bm\zeta]=c_0+c_1|{\bm f}|^2,
\label{eq:E1}
\end{align}
where
\begin{align}
 f_+&\equiv f_x+if_y
 =\sqrt{2}\zeta_1^*\zeta_0+\sqrt{2}\zeta_0^*\zeta_{-1},\\
 f_z&=|\zeta_1|^2-|\zeta_{-1}|^2.
\end{align}

\subsubsection{Continuous symmetry}
There are two inert states that have continuous isotropy groups:
\begin{align}
F:&\ \bm\zeta^{(1)}=(1,0,0)^T,\label{eq:spin1F}\\
P:&\ \bm\zeta^{(0)}=(0,1,0)^T,\label{eq:spin1P}
\end{align}
where the former is the ferromagnetic state, while the latter is the polar (or antiferromagnetic) state.
The isotropy group of these states are given by substituting $(F,m_0)=(1,1)$ and $(1,0)$ to Eqs.~\eqref{eq:H^m_0} and \eqref{eq:H^m_00}, respectively, as
\begin{align}
F:&\ \ H^{(1)} = \{e^{-iF_z\alpha} e^{i\alpha}, e^{i2\gamma}U_2^\gamma \mathcal{T}\},\\
P:&\ \ H^{(0)} = \{e^{-iF_z\alpha} , e^{i\pi}U_2^\gamma \}\times \Theta,
\end{align}
where $\alpha$ and $\gamma$ are arbitrary real numbers.
Substituting $\bm\zeta^{(1,0)}$ in Eq.~\eqref{eq:E1},
the mean-field energies are obtained as 
\begin{align}
F:&\ \ \mathcal{E}^{(1)}[\bm\zeta^{(1)}]=c_0+c_1,\\
P:&\ \ \mathcal{E}^{(1)}[\bm\zeta^{(0)}]=c_0.
\end{align}

\subsubsection{Discrete symmetry}
\begin{figure}[ht]
\includegraphics[scale=0.35]{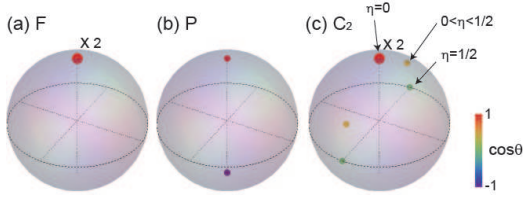}
\caption{(Color online) Majorana representation of (a) ferromagnetic state $\bm\zeta^{(1)}$, (b) polar state $\bm\zeta^{(0)}$, and (c) $\bm\zeta^{(C_2)}$ state [Eq.~\eqref{eq:spin1_C2}] with $C_2$ symmetry.
The color of the points shows the value of $\cos\theta=(1-|z|^2)/(1+|z|^2)$ according to the color gauge.
For the case of the ferromagnetic state, the polynomial $P_{\bm \zeta}^{(1)}(z)=z^2$ has two degenerate roots at the north pole, which is indicated with $\times 2$ in (a).
In (c), two vertices are on the $y$--$z$ plane on the same latitude.
They are both at the north pole for $\eta=0$ and on the equator for $\eta=1/2$.
In (c) the $\bm\zeta^{(C_2)}(\eta=1/2)$ state  has the same symmetry as the polar state.
}
\label{fig:spin1}
\end{figure}

We first consider the eigenstate of $C_{nz}$ whose matrix representation in the spin-1 manifold is given by
\begin{align}
C_{nz} = {\rm Diag}[e^{-i 2\pi /n}, 1, e^{i 2\pi /n}],
\end{align}
where ${\rm Diag}$ means the diagonal matrix.
A nontrivial eigenstate of $C_{nz}$ that has more than two non-zero components exists only for $n=2$ and is given by
\begin{align}
 C_2:&\ \bm\zeta^{(C_2)}=(\sqrt{1-\eta}, 0, \sqrt{\eta})^T
\label{eq:spin1_C2}
\end{align}
with eigenvalue $-1$, where $0 < \eta \le 1/2$.
Note that we can always choose $\zeta_1$ and $\zeta_{-1}$ to be real and positive without loss of generality,
since the phase factors of these components can be removed by a spin rotation about the $z$ axis and a gauge transformation.
The geometric structures of $\bm\zeta^{(1)}$, $\bm\zeta^{(0)}$, and $\bm\zeta^{(C_2)}$ are shown in Fig.~\ref{fig:spin1}.

For the case of $H'=D_2$, the order parameter should be an eigenstate of $C_{2x}$:
\begin{align}
 C_{2x} = \begin{pmatrix} 0 & 0 & -1 \\ 0 & -1 & 0 \\ -1 & 0 & 0 \end{pmatrix},
\end{align}
resulting in $\eta=1/2$.
This state is nothing but the polar state since $\bm\zeta^{(C_2)}(\eta=1/2)=(1/\sqrt{2},0,1/\sqrt{2})^T$ is related to $\bm\zeta^{(0)}$ by rotation: $e^{i\pi/2}e^{-iF_y\pi/2}\bm\zeta^{(0)}=\bm\zeta^{(C_2)}(\eta=1/2)$.
In other words, $\bm\zeta^{(0)}$ and $\bm\zeta^{(C_2)}(\eta=1/2)$ are on the same orbit.
This fact can also be understood by comparing the geometric structures of these two states (Fig.~\ref{fig:spin1}).
For the case of $H'=C_2$, 
substituting the order parameter~\eqref{eq:spin1_C2} in Eq.~\eqref{eq:E1},
we obtain
\begin{align}
\mathcal{E}^{(1)}[\bm\zeta^{(C_2)}] = c_0 + c_1 (1-2\eta)^2.
\end{align}
The stationary point of $\mathcal{E}^{(1)}$ is at $\eta=1/2$, and the corresponding state has the same symmetry as the polar state.
Hence, there are only two phases in a spin-1 BEC: the ferromagnetic phase and the polar phase given by Eqs.~\eqref{eq:spin1F} and \eqref{eq:spin1P}, respectively.

\subsubsection{Phase Diagram}
\begin{figure}[ht]
\includegraphics[scale=0.5]{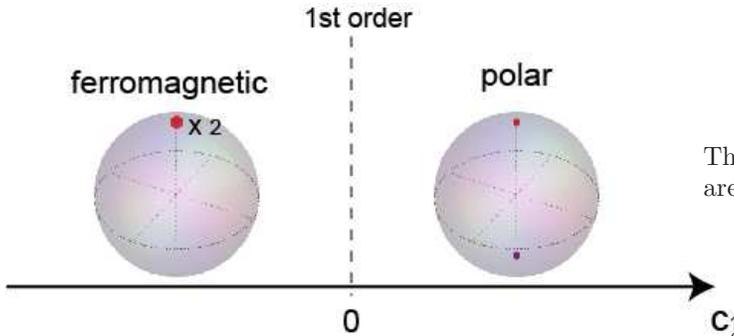}
\caption{(Color online) Phase diagram of a spin-1 BEC. The phase transition at $c_1=0$ is first-order
because the symmetry changes discontinuously at that point.
}
\label{fig:spin1PD}
\end{figure}

Comparing the energies of the polar and ferromagnetic phases, we obtain the phase diagram of the spin-1 system as shown in Fig.~\ref{fig:spin1PD}.
The physics of the phase diagram is quite simple:
from the mean-field energy~\eqref{eq:E1}, $|\bm f|$ should vanish for $c_1>0$, whereas it becomes maximal (i.e., $|\bm f|=1$) for $c_1<0$;
the former is polar and the latter is ferromagnetic~\cite{Ohmi1998,Ho1998}.

\subsection{Spin-2}
\label{sec:spin2}

\subsubsection{Mean-field energy}
A spin-2 spinor BEC is described with a five-component spinor $\bm\zeta=(\zeta_2,\zeta_1,\zeta_0,\zeta_{-1},\zeta_{-2})^T$.
The scaled mean-field energy for a given order parameter $\bm\zeta$ is written as
\begin{align}
\mathcal{E}^{(2)}[\bm\zeta]=c_0+c_1|{\bm f}|^2+c_2|A_{00}|^2,
\label{eq:E2}
\end{align}
where
\begin{align}
 f_+&=2\zeta_2^*\zeta_1+\sqrt{6}\zeta_1^*\zeta_0+\sqrt{6}\zeta_0^*\zeta_{-1}+2\zeta_{-1}^*\zeta_{-2},\\
 f_z&=2|\zeta_2|^2+|\zeta_1|^2-|\zeta_{-1}|^2-2|\zeta_{-2}|^2,\\
 A_{00} &= \frac{1}{\sqrt{5}}(2\zeta_2\zeta_{-2}-2\zeta_1\zeta_{-1}+\zeta_0^2).
\end{align}

\subsubsection{Continuous symmetry}
\begin{figure}[ht]
\includegraphics[scale=0.37]{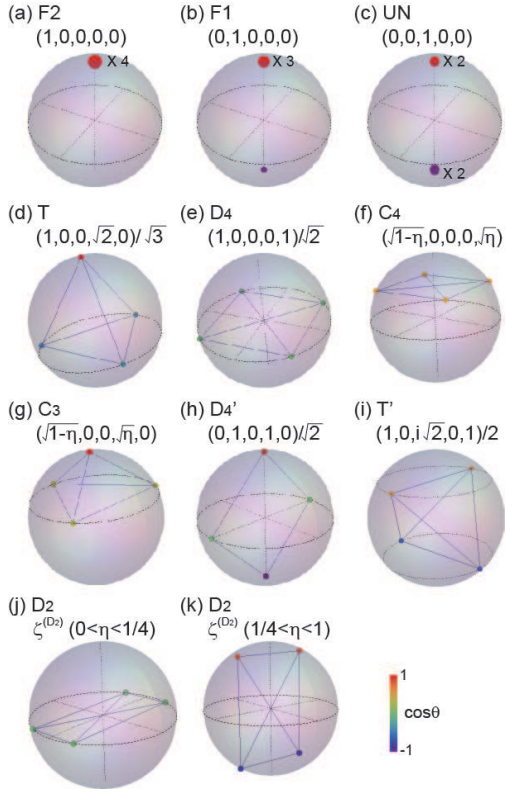}
\caption{(Color online) Majorana representation of spin-2 BECs.
(a) $F2$, (b) $F1$, and (c) uniaxial nematic ($UN$) states. The $F2$ and $F1$ states have the $SO(2)$ symmetry, 
whereas the symmetry group of the $UN$ state is $D_\infty$.
(d) Cyclic state with the tetrahedron symmetry.
Four vertices in (d) form a regular tetrahedron.
(e) $D_4$ state with four vertices forming a square.
(f) State with the $C_4$ symmetry given in Eq.~\eqref{eq:spin2C4}.
Depending on $\eta$,
the square moves between the north pole ($\eta=0$) and the equator ($\eta=1/2$).
(g) State with the $C_3$ symmetry given in Eq.~\eqref{eq:spin2C3}.
The triangle in (g) moves between the north pole ($\eta=0$) and the south pole ($\eta=1$),
with one vertex fixed at the north pole.
(h) and (i) represent the states $(0,1,0,1,0)^T/\sqrt{2}$ and $(1,0,i\sqrt{2},0,1)^T/2$, respectively.
(h) has the same symmetry as (e), and (i) has the same symmetry as (d).
(j) $\bm\zeta^{(D_2)}=(\sqrt{(1-\eta)/2},0,\sqrt{\eta},0,\sqrt{(1-\eta)/2})^T$ for $0<\eta<1/4$
with four vertices forming a rectangle on the equator.
(k) This state has the same order parameter as (j) but for $1/4<\eta<1$.
The state $\bm\zeta^{(D_2)}$ has the same symmetry as the $D_4$ state when $\eta=0$ and $1/2$, and as the {\it UN} state when $\eta=1/4$ and $1$.
}
\label{fig:spin2}
\end{figure}

There are three inert states that have continuous isotropy groups:
\begin{align}
F2:&\ \bm\zeta^{(2)}=(1,0,0,0,0)^T, \label{eq:zeta_spin2F2}\\
F1:&\ \bm\zeta^{(1)}=(0,1,0,0,0)^T, \label{eq:zeta_spin2F1}\\
UN:&\ \bm\zeta^{(0)}=(0,0,1,0,0)^T, \label{eq:zeta_spin2UN}
\end{align}
where $F$ and $UN$ stand for ferromagnetic and uniaxial-nematic~\cite{Song2007,Turner2007}, respectively.
The isotropy groups of these states are obtained by substituting $(F,m_0)=(2,2), (2,1)$ and $(2,0)$ in Eqs.~\eqref{eq:H^m_0} and \eqref{eq:H^m_00}, respectively:
\begin{align}
F2:&\ \ H^{(2)} = \{ e^{2i\alpha} e^{-iF_z\alpha}, e^{4i\gamma}U_2^\gamma \mathcal{T}\},\label{eq:gene_spin2F2}\\
F1:&\ \ H^{(1)} = \{ e^{i\alpha} e^{-iF_z\alpha} , e^{i(2\gamma+\pi)}U_2^\gamma \mathcal{T}\},\label{eq:gene_spin2F1}\\
UN:&\ \ H^{(0)} = \{e^{-iF_z\alpha} , U_2^\gamma\}\times \Theta,\label{eq:gene_spin2UN}
\end{align}
where $\alpha$ and $\gamma$ are arbitrary real numbers.
Substituting $\bm\zeta^{(2,1,0)}$ in Eq.~\eqref{eq:E2}, we obtain
\begin{align}
F2:&\ \ \mathcal{E}^{(2)}[\bm\zeta^{(2)}]=c_0+4c_1,\label{eq:ene_spin2F2}\\
F1:&\ \ \mathcal{E}^{(2)}[\bm\zeta^{(1)}]=c_0+c_1,\label{eq:ene_spin2F1}\\
UN:&\ \ \mathcal{E}^{(2)}[\bm\zeta^{(0)}]=c_0+\frac{c_2}{5}.\label{eq:ene_spin2UN}
\end{align}
The Majorana representations of $F2$, $F1$, and $UN$ states
are shown in Figs.~\ref{fig:spin2}(a), \ref{fig:spin2}(b) and \ref{fig:spin2}(c), respectively.

\subsubsection{Discrete symmetry}
Since a spin-2 system is described with four vertices in the Majorana representation,
the symmetries that a spin-2 BEC may have are $T$, $D_{4,3,2}$, and $C_{4,3,2}$.
For each symmetry group, we seek stationary points of $\mathcal{E}^{(2)}[\bm\zeta]$.
The Majorana representation of the obtained states are shown in Fig.~\ref{fig:spin2} (see also Ref.~\cite{Turner2007}).

\begin{itemize}
\item [$T$:]
The generators of the tetrahedron group are $C_{3z}$ and $C_{2,\sqrt{2}x+z}$ whose matrix representation 
on the spin-2 manifold are given by
\begin{align}
&C_{3z} = {\rm Diag}[e^{i2\pi/3}, e^{i4\pi/3}, 1, e^{i2\pi/3}, e^{i4\pi/3}],\label{eq:spin2_matrixC3z}\\
& C_{2,\sqrt{2}x+z} = \exp\left(-i\frac{\sqrt{2}F_x+F_z}{\sqrt{3}}\pi\right)\\
&=\frac{1}{9}
\begin{pmatrix}
        1 &  2\sqrt{2} &  2\sqrt{6} &  4\sqrt{2} &          4 \\
2\sqrt{2} &          5 &  2\sqrt{3} &         -2 & -4\sqrt{2} \\
2\sqrt{6} &  2\sqrt{3} &         -3 & -2\sqrt{3} &  2\sqrt{6} \\
4\sqrt{2} &         -2 & -2\sqrt{3} &          5 & -2\sqrt{2} \\
        4 & -4\sqrt{2} &  2\sqrt{6} & -2\sqrt{2} &          1
\end{pmatrix}.
\end{align}
The simultaneous eigenstate of these two operators is uniquely determined to be
\begin{align}
T:\ \bm\zeta^{(T)}=(1/\sqrt{3},0,0,\sqrt{2/3},0)^T,
\label{eq:spin2_OPcyclic}
\end{align}
or its time reversal $\mathcal{T}\bm\zeta^{(T)}=(0,-\sqrt{3/2},0,0,1/\sqrt{3})^T$,
up to an overall gauge. The eigenvalues of $\bm \zeta^{(T)}$ 
are $e^{i2\pi/3}$ for $C_{3z}$ and 1 for $C_{2,\sqrt{2}x+z}$.
This state is called the cyclic state.
The isotropy group is generated by a set of the following operators:
\begin{align}
&\tilde{H}^{(T)} = \{e^{-i2\pi/3}C_{3z}, C_{2,\sqrt{2}x+z}, e^{-iF_z\gamma}U_2^\gamma\mathcal{T}\},
\end{align}
where $\gamma$ is an arbitrary real number, and here and henceforth, we denote a set of generators of $H$ by $\tilde{H}$.
Though no spontaneous magnetization arises in the cyclic state,
the time reversal symmetry is broken, because $\mathcal{T}\bm\zeta^{(T)} \neq \bm\zeta^{(T)}$.
One can also confirm this fact from the Majorana representation as shown in Fig.~\ref{fig:spin2}(d):
the antipoles of the vertices in Fig.~\ref{fig:spin2}(d) form a time-reversed tetrahedron, but it does not coincide with the original one.
The mean-field energy of the cyclic state is given by
\begin{align}
\mathcal{E}^{(2)}[\bm\zeta^{(T)}] = c_0.
\label{eq:ene_spin2T}
\end{align}

\item [$D_4$:]
A nontrivial eigenstate of 
\begin{align}
C_{4z}={\rm Diag}[-1, -i, 1, i, -1]
\end{align}
is written without loss of generality as
\begin{align}
(\sqrt{1-\eta},0,0,0,\sqrt{\eta})^T,
\label{eq:spin2C4}
\end{align}
where $0 < \eta\le 1/2$.
When $\eta=1/2$, this state becomes an eigenstate of $C_{2x}$ with eigenvalue 1.
Hence, there is an inert state that has the $D_4$ symmetry:
\begin{align}
D_4:\ \bm\zeta^{(D_4)}=(1/\sqrt{2},0,0,0,1/\sqrt{2})^T.
\label{eq:spin2BN}
\end{align}
This state is often refereed to as the biaxial nematic state~\cite{Song2007,Turner2007}.
The generators of the isotropy group and the mean-field energy for this state are given by
\begin{align}
&\tilde{H}^{(D_4)} = \{e^{-i\pi}C_{4z}, C_{2x}, \mathcal{T}\},\\
&\mathcal{E}^{(2)}[\bm\zeta^{(D_4)}] = c_0 +\frac{c_2}{5},\label{eq:ene_spin2BN}
\end{align}
respectively.
Though the energy of $\bm\zeta^{(D_4)}$ is the same as that of $\bm\zeta^{(0)}$,
the geometric structures of these states are different from each other as shown in Figs.~\ref{fig:spin2}(c) and \ref{fig:spin2}(e).

\item [$C_4$:]
The energy of the eigenstate of $C_{4z}$ [Eq.~\eqref{eq:spin2C4}] is calculated as a function of $\eta$ as
\begin{align}
\mathcal{E}^{(2)} &= c_0 + 4c_1(1-2\eta)^2 + \frac{4c_2}{5}\eta(1-\eta).\label{eq:spin2_energy_C4}
\end{align}
Equation~\eqref{eq:spin2_energy_C4} has a stationary point at $\eta=1/2$, resulting in the $D_4$ state.
The $\eta$ dependence of the geometric structure of the state given in Eq.~\eqref{eq:spin2C4} is shown in Fig.~\ref{fig:spin2}(f).

\item [$D_3$:]
The eigenstate of $C_{3z}$ is given by
\begin{align}
(\sqrt{1-\eta},0,0,\sqrt{\eta},0)^T,
\label{eq:spin2C3}
\end{align}
where $0 < \eta < 1$.
The order parameter in the form of $(0,\sqrt{\eta},0,0,\sqrt{1-\eta})^T$ is also a nontrivial eigenstate of $C_{3z}$.
However, this state belongs to the same orbit of that in Eq.~\eqref{eq:spin2C3}, because they are transformed into each other by a spin rotation and a gauge transformation.
There is no simultaneous eigenstate of $C_{3z}$ and $C_{2x}$,
i.e., there is no state that has the $D_3$ symmetry.

\item[$C_3$:]
The energy of the eigenstate of $C_3$ [Eq.~\eqref{eq:spin2C3}] is calculated as a function of $\eta$ as
\begin{align}
\mathcal{E}^{(2)} &= c_0 + c_1(2-3\eta)^2.\label{eq:spin2_energy_C3}
\end{align}
The stationary point of this function is $\eta=2/3$, and the corresponding order parameter is identical to the cyclic state [Eq.~\eqref{eq:spin2_OPcyclic}].
The geometric structure of the state given in Eq.~\eqref{eq:spin2C3} is shown in Fig.~\ref{fig:spin2}(g).

\item [$D_2$:]
The matrix representation of $C_{2z}$ is given by
\begin{align}
C_{2z}={\rm Diag}[1,-1,1,-1,1]
\end{align}
There are two simultaneous eigenstates of $C_{2z}$ and $C_{2x}$.

{\it Case} (i): The order parameter
\begin{align}
(0,1/\sqrt{2},0,1/\sqrt{2},0)^T
\end{align}
is the simultaneous eigenstates of $C_{2z}$ and $C_{2x}$ with eigenvalues $-1$ and $1$, respectively.
However, this state has the same symmetry as that of the biaxial nematic ($D_4$) state as shown in Fig.~\ref{fig:spin2}(h).

{\it Case} (ii): The other simultaneous eigenstate can be written as
\begin{align}
\left(\sqrt{\frac{1-\eta}{2}},0,e^{i\delta}\sqrt{\eta},0,\sqrt{\frac{1-\eta}{2}}\right)^T,
\end{align}
where the eigenvalues of $C_{2z}$ and $C_{2x}$ are both equal to $1$.
The mean-field energy of this state is given as a function of $\eta$ and $\delta$ as
\begin{align}
&\mathcal{E}^{(2)}= c_0 + \frac{c_2}{5}[1-2\eta+2\eta^2+2\eta(1-\eta)\cos2\delta].\label{eq:spin2_energy_D2'}
\end{align}
Taking the partial derivatives of Eq.~\eqref{eq:spin2_energy_D2'} with respect to $\delta$ and $\eta$, we obtain 
two stationary points at $\delta=\pi/2$ and $\eta=1/2$, and at $\delta=0$ and arbitrary $\eta$.
For the former case, the corresponding order parameter, $(1/2,0,i/\sqrt{2},0,1/2)$, has the tetrahedral symmetry as shown in Fig.~\ref{fig:spin2}(i).
For the latter case, the order parameter is given by
\begin{align}
D_2:\ \bm\zeta^{(D_2)}=\left(\sqrt{\frac{1-\eta}{2}}, 0, \sqrt{\eta}, 0, \sqrt{\frac{1-\eta}{2}}\right)^T,
\label{eq:spin2D2}
\end{align}
which has symmetry different from that of the other obtained state [Fig.~\ref{fig:spin2}(i)], as shown in Figs.~\ref{fig:spin2}(j) and \ref{fig:spin2}(k).
The energy for this state is calculated to be
\begin{align}
\mathcal{E}^{(2)}[\bm\zeta^{(D_2)}]=c_0+\frac{c_2}{5},\label{eq:ene_spin2D2}
\end{align}
which does not depend on $\eta$,
implying that all the states described by Eq.~\eqref{eq:spin2D2} are degenerate.
The uniaxial and biaxial nematic states are also included in Eq.~\eqref{eq:spin2D2},
and they can be smoothly transformed to each other by changing $\eta$ in Eq.~\eqref{eq:spin2D2}.
For a fixed $\eta$, the generators of the isotropy group of $\bm\zeta^{(D_2)}$ is given by
\begin{align}
\tilde{H}^{(D_2)} =\{C_{2z},C_{2x},\mathcal{T}\}.
\end{align}
However, if we take into account the degrees of freedom described by $\eta$, the isotropy group of the state given in Eq.~\eqref{eq:spin2D2} is shown to be $[\mathbb{Z}_2 \rtimes SO(4)]\times \Theta$~\cite{Uchino2010b},
where $\rtimes$ implies that the nontrivial element of $\mathbb{Z}_2$ does not commute with some elements of $SO(4)$.
It has also been pointed out that the degeneracy with respect to $\eta$ is lifted if we take into account quantum or thermal fluctuations~\cite{Song2007,Turner2007,Uchino2010a}.

\item [$C_2$:]
There are two nontrivial eigenstates of $C_{2z}$.

{\it Case} (i): The order parameter
\begin{align}
(0,\sqrt{1-\eta},0,\sqrt{\eta},0)^T,\label{eq:spin2C2}
\end{align}
is the eigenstate of $C_{2z}$ with eigenvalue $-1$.
The mean-field energy of this state is given by
\begin{align}
\mathcal{E}^{(2)} &= c_0 + c_1(1-2\eta)^2 + \frac{4c_2}{5}\eta(1-\eta).\label{eq:spin2_energy_C2}
\end{align}
This function has a stationary point at $\eta=1/2$.
The corresponding state has the same symmetry as the biaxial nematic ($D_4$) state.

{\it Case} (ii): The order parameter
\begin{align}
(a_+,0,b,0,a_-)^T
\end{align}
is the eigenstate of $C_{2z}$ with eigenvalue 1,
where $a_\pm$ and $b$ are complex numbers.
Here, we choose $a_\pm$ to be real numbers and rewrite these parameters as
\begin{align}
a_\pm&=\sqrt{\frac{1-\eta-\xi}{2}}\pm\sqrt{\frac{\xi}{2}},\\
b&=e^{i\delta}\sqrt{\eta},
\end{align}
where $\eta,\xi\ge 0$, $\eta+\xi\le1$, and $-\pi < \delta\le \pi$.
The mean-field energy of this state is given by
\begin{align}
\mathcal{E}^{(2)}=&c_0+16 c_1\xi(1-\eta-\xi)\nonumber\\
&+\frac{c_2}{5}|1-2\xi-(1-e^{2i\delta})\eta|^2.\label{eq:spin2_energy_C2'}
\end{align}
Taking the partial derivatives of Eq.~\eqref{eq:spin2_energy_C2'} with respect to $\delta$, $\eta$, and $\xi$, we obtain
the same stationary solutions as those in case (ii) of the $D_2$ symmetry.
\end{itemize}

\subsubsection{Phase diagram}
\label{sec:spin2PD}
\begin{figure}[htb]
\includegraphics[scale=0.4]{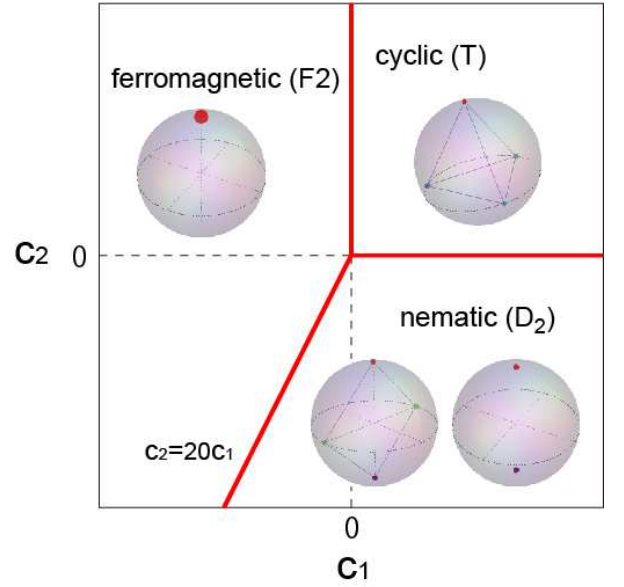}
\caption{(Color online) Phase diagram of a spin-2 BEC. 
In the nematic (antiferromagnetic) phase, all states described by Eq.~\eqref{eq:spin2D2} are degenerate. 
The phase boundaries on $c_1=0, c_2=0$ and $c_2=c_1/20$ are all first-order
because the symmetry changes discontinuously across the boundaries.
}
\label{fig:spin2PD}
\end{figure}

Comparing the energies of the obtained stationary solutions [Eqs.~\eqref{eq:ene_spin2F2}, \eqref{eq:ene_spin2F1}, \eqref{eq:ene_spin2T}, and \eqref{eq:ene_spin2D2}],
we obtain the phase diagram of a spin-2 BEC as shown in Fig.~\ref{fig:spin2PD}.
In the region of nematic phase of Fig.~\ref{fig:spin2PD}, all states described by Eq.~\eqref{eq:spin2D2} are degenerate, including uniaxial and biaxial nematic states.
Our results agree well with those in the previous works~\cite{Koashi2000,Ciobanu2000,Ueda2002}.
Distinct from the case for $F=1$, the phase diagram for $F=2$ is determined by the last two terms in Eq.~\eqref{eq:E2}: $c_1|\bm f|^2$ and $c_2|A_{00}|^2$.
Clearly, $|\bm f|$ can vary within $0\le |\bm f|\le 2$ for an $F=2$ system.
Note that $|A_{00}|$ is proportional to the inner product of the order parameter and its time reversal: $|A_{00}|=(\mathcal{T}\bm\zeta)^\dagger \bm\zeta/\sqrt{5}$.
It takes the maximum value of $1/\sqrt{5}$ when the order parameter has the time reversal symmetry, while it should vanish for the ferromagnetic state.
Then, the ferromagnetic phase arises for $c_1<0$ and $c_2> 20c_1$, while the nematic state, which has the time reversal symmetry, becomes the ground state for $c_2<0$ and $c_1>c_2/20$.
In the region of $c_1>0$ and $c_2>0$, the cyclic phase appears since both $|\bm f|$ and $|A_{00}|$ vanish in this phase.

\subsection{Spin-3}
\label{sec:spin3}

\subsubsection{Mean-field energy}
A spin-3 BEC is described with a seven-component order parameter $\bm\zeta=(\zeta_3,\zeta_2,\zeta_1,\zeta_0,\zeta_{-1},\zeta_{-2},\zeta_{-3})^T$.
The scaled mean-field energy is given by Eq.~\eqref{eq:E3}.
Following the notations in Ref.~\cite{Diener2006},
we rewrite Eq.~\eqref{eq:E3} in terms of $B_{00}\equiv \sqrt{7}A_{00}$ and $B_{2M}=\sqrt{7}A_{2M}$ as
\begin{align}
\tilde{\mathcal{E}}^{(3)}[\bm\zeta] \equiv& \mathcal{E}^{(3)}[\bm\zeta] -c_0 \nonumber\\
=& c_\gamma |\bm f|^2 + c_\alpha |B_{00}|^2 + c_\beta \sum_{M=-2}^2|B_{2M}|^2,
\label{eq:scaledE3}
\end{align}
where $c_\gamma=c_1$, $c_\alpha=c_2/7$, $c_\beta=c_3/7$, which correspond to $\gamma$, $\alpha$, and $\beta$ in Ref.~\cite{Diener2006}, respectively,
\begin{align}
f_+=&\sqrt{6}\zeta^*_{3}\zeta_{2}+\sqrt{10}\zeta^*_{2}\zeta_{1}+2\sqrt{3}\zeta^*_{1}\zeta_{0}\nonumber\\
&+2\sqrt{3}\zeta^*_{0}\zeta_{-1}+\sqrt{10}\zeta^*_{-1}\zeta_{-2}+\sqrt{6}\zeta^*_{-2}\zeta_{-3}, \label{eq:spin3_fpm}\\
f_z=&3|\zeta_3|^2+2|\zeta_2|^2+|\zeta_1|^2-|\zeta_{-1}|^2-2|\zeta_{-2}|^2-3|\zeta_{-3}|^2,\label{eq:spin3_fz}
\end{align}
are the spin densities, and
\begin{align}
B_{00}=&2\zeta_3\zeta_{-3}-2\zeta_2\zeta_{-2}+2\zeta_{1}\zeta_{-1}-\zeta_{0}^2,\\
B_{2 0}=& \frac{5}{\sqrt{3}}\zeta_3\zeta_{-3}   - \sqrt{3}\zeta_1\zeta_{-1}            +\sqrt{\frac{2}{3}}\zeta_0^2,\\
B_{2,\pm1}=& \frac{5}{\sqrt{3}}\zeta_{\pm3}\zeta_{\mp2} - \sqrt{5}\zeta_{\pm2}\zeta_{\mp1}          +\sqrt{\frac{2}{3}}\zeta_{\pm1}\zeta_0,\\
B_{2,\pm2}=& \sqrt{\frac{10}{3}}\zeta_{\pm3}\zeta_{\mp1} - \sqrt{\frac{20}{3}}\zeta_{\pm2}\zeta_0 +\sqrt{2}\zeta_{\pm1}^2.
\end{align}

\subsubsection{Continuous symmetry}
There are four inert states that have continuous isotropy groups:
\begin{align}
&\ \bm\zeta^{(3)}=(1,0,0,0,0,0,0),\\
&\ \bm\zeta^{(2)}=(0,1,0,0,0,0,0),\\
&\ \bm\zeta^{(1)}=(0,0,1,0,0,0,0),\\
&\ \bm\zeta^{(0)}=(0,0,0,1,0,0,0).\label{eq:spin3P0}
\end{align}
The isotropy groups of these states are given by substituting $(F,m_0)=(3,3), (3,2)$ and $(3,1)$ in Eq.~\eqref{eq:H^m_0}, and $(F,m_0)=(3,0)$ in Eq.~\eqref{eq:H^m_00} as
\begin{align}
&\ \ H^{(3)} = \{e^{-iF_z\alpha} e^{3i\alpha}, U_2^\gamma e^{i6\gamma}\mathcal{T}\},\label{eq:Hspin3F3}\\
&\ \ H^{(2)} = \{e^{-iF_z\alpha} e^{2i\alpha}, U_2^\gamma e^{i(4\gamma+\pi)}\mathcal{T}\},\label{eq:Hspin3F2}\\
&\ \ H^{(1)} = \{e^{-iF_z\alpha} e^{i\alpha}, U_2^\gamma e^{i2\gamma}\mathcal{T}\},\label{eq:Hspin3F1}\\
&\ \ H^{(0)} = \{e^{-iF_z\alpha} , U_2^\gamma e^{i\pi}\} \times \Theta,\label{eq:Hspin3P0}
\end{align}
where $\alpha$ and $\gamma$ are arbitrary real numbers.
Substituting $\bm\zeta^{(3, 2,1,0)}$ in Eq.~\eqref{eq:scaledE3},
the mean-field energies are obtained as 
\begin{align}
&\ \ \tilde{\mathcal{E}}^{(3)}[\bm\zeta^{(3)}]=9c_\gamma,\label{eq:ene_spin3F3}\\
&\ \ \tilde{\mathcal{E}}^{(3)}[\bm\zeta^{(2)}]=4c_\gamma,\label{eq:ene_spin3F2}\\
&\ \ \tilde{\mathcal{E}}^{(3)}[\bm\zeta^{(1)}]=c_\gamma+2c_\beta,\label{eq:ene_spin3F1}\\
&\ \ \tilde{\mathcal{E}}^{(3)}[\bm\zeta^{(0)}]=c_\alpha+\frac{4c_\beta}{3}.\label{eq:ene_spin3P0}
\end{align}

\subsubsection{Discrete symmetry}
Since the state of the spin-3 BEC is represented by six vertices in the Majorana representation,
the possible symmetries for a spin-3 BEC are $O$, $T$, $D_{6,5,4,3,2}$, and $C_{6,5,4,3,2}$.
For each symmetry group, we seek stationary points of $\tilde{\mathcal{E}}^{(3)}[\bm\zeta]$.

\begin{itemize}
\item[$O$:]
The generators of the octahedron symmetry are $C_{4z}$ and $C_{2,x+z}$
whose representations on a spin-3 manifold are given by
\begin{widetext}
\begin{align}
&C_{4z}={\rm Diag}[i,-1,-i,1,i,-1,-i],\label{eq:spin3_matrixC4z}\\
&C_{2,x+z} = \exp\left(-i\frac{F_x+F_z}{\sqrt{2}}\pi\right)
=\frac{1}{8}\begin{pmatrix}
-1& -\sqrt{6}& -\sqrt{15}& -2 \sqrt{5}& -\sqrt{15}& -\sqrt{6}& -1\\
-\sqrt{6}& -4& -\sqrt{10}& 0& \sqrt{10}& 4& \sqrt{6}\\
-\sqrt{15}& -\sqrt{10}& 1& 2 \sqrt{3}& 1& -\sqrt{10}& -\sqrt{15}\\
-2 \sqrt{5}& 0& 2 \sqrt{3}& 0& -2 \sqrt{3}& 0& 2 \sqrt{5}\\
-\sqrt{15}& \sqrt{10}& 1& -2 \sqrt{3}& 1& \sqrt{10}& -\sqrt{15}\\
-\sqrt{6}& 4& -\sqrt{10}& 0& \sqrt{10}& -4& \sqrt{6}\\
-1& \sqrt{6}& -\sqrt{15}& 2 \sqrt{5}& -\sqrt{15}& \sqrt{6}& -1
  \end{pmatrix},
\end{align}
\end{widetext}
respectively.
The simultaneous eigenstate of these operators is uniquely obtained as
\begin{align}
O:\ \bm\zeta^{(O)}=(0,1/\sqrt{2},0,0,0,-1/\sqrt{2},0)^T.
\label{eq:spin3O}
\end{align}
Here, the corresponding eigenvalues are equal to $-1$ for both $C_{4z}$ and $C_{2,x+z}$.
Then, the generators of the isotropy group are given by
\begin{align}
\tilde{H}^{(O)}=\{e^{i\pi}C_{4z}, e^{i\pi}C_{2,x+z}, e^{i\pi}\mathcal{T}\}.
\end{align}
The mean-field energy is calculated as
\begin{align}
\tilde{\mathcal{E}}^{(3)}[\bm\zeta^{(O)}]=c_\alpha.
\end{align}

\item [$T$:]
The generators of the tetrahedron symmetry are $C_{3z}$ and $C_{2,\sqrt{2}x+z}$ whose matrix representations are given by
\begin{widetext}
\begin{align} 
&C_{3z}={\rm Diag}[1,e^{i2\pi/3},e^{i4\pi/3},1,e^{i2\pi/3},e^{i4\pi/3},1],\\
&C_{2,\sqrt{2}x+z} = \exp\left(-i\frac{\sqrt{2}F_x+F_{z}}{\sqrt{3}}\pi\right)
=\frac{1}{27}
\begin{pmatrix}
-1         & -2\sqrt{3} & -2\sqrt{15}& -4\sqrt{10}& -4\sqrt{15}& -8\sqrt{3} & -8\\
-2\sqrt{3} & -9         & -6\sqrt{5} & -2\sqrt{30}& 0          & 12         & 8\sqrt{3}\\
-2\sqrt{15}& -6\sqrt{5} & -9         & 2\sqrt{6}  & 12         & 0          & -4\sqrt{15}\\
-4\sqrt{10}& -2\sqrt{30}& 2\sqrt{6}  & 11         & -2\sqrt{6} & -2\sqrt{30}& 4\sqrt{10}\\
-4\sqrt{15}& 0          & 12         & -2\sqrt{6} & -9         & 6\sqrt{5}  & -2\sqrt{15}\\
-8\sqrt{3} & 12         & 0          & -2\sqrt{30}& 6\sqrt{5}  & -9         & 2\sqrt{3}\\
-8         & 8\sqrt{3}  & -4\sqrt{15}& 4\sqrt{10} & -2\sqrt{15}& 2\sqrt{3}  & -1
\end{pmatrix}.
\end{align}
\end{widetext}
These two operators have a unique simultaneous eigenstate
\begin{align}
 (\sqrt{2}/3,0,0,-\sqrt{5}/3,0,0,-\sqrt{2}/3).
\label{eq:spin3T}
\end{align}
However, this state has the same symmetry as the $O$ state.
In fact, the Majorana representation of Eq.~\eqref{eq:spin3T} is obtained by rotating that the state in Eq.~\eqref{eq:spin3O}
\end{itemize}

\begin{figure}[ht]
\includegraphics[scale=0.4]{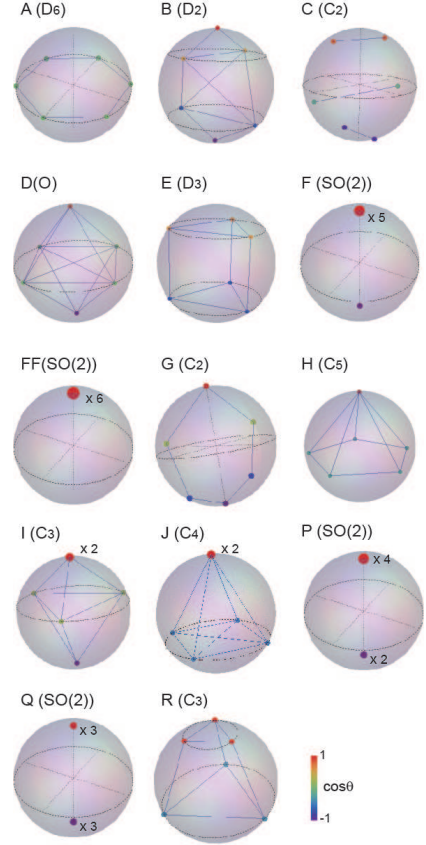}
\caption{(Color online) Majorana representation of stationary states of spin-3 BECs. 
The order parameter for each state is listed in Table~\ref{table:spin3}.
}
\label{fig:spin3-1}
\end{figure}

For the symmetry groups of $D_n$ and $C_n$, we proceed in a manner  similar to the case of the spin-2 BEC.
We characterize the eigenstate of $C_{nz}$ with a few parameters ($\eta, \xi$, etc.).
For the $D_6$ symmetry, the order parameter is uniquely determined; therefore, this is an inert state.
For the other cases, we rewrite $\tilde{\mathcal{E}}^{(3)}$ in terms of the new parameters
and find stationary points.
The detailed calculations are described in Appendix~\ref{sec:append_spin3}.
The results are summarized in Table~\ref{table:spin3}, in which we list all the obtained stationary states, together with their isotropy groups.
We have obtained the analytical solutions for all stationary states, except for state C.
For state C we have numerically calculated the energy by restricting the order parameter in the form of $(a,0,b,0,c,0,d)^T$ with $a,b,c,d\in\mathbb{C}$.
In the obtained state, $a,b$ and $d$ are real positive numbers and $c$ is a real negative number.
The geometric structure of the obtained states are shown in Fig.~\ref{fig:spin3-1} (see also Ref.~\cite{Barnett2007}).

Among the obtained stationary states, only states A, D, and Q possess the time-reversal symmetry.
The time-reversal operator $\mathcal{T}$ is decoupled from spin rotations,
which can also be understood from Fig.~\ref{fig:spin3-1}, where
the antipodal map does not change the configurations of vertices for the A, D, and Q states.
The time-reversal operation changes the configurations of vertices for other states.
In particular, the time-reversal symmetry is broken in the B and E states, even though these states have no spontaneous magnetization,
as in the case of the cyclic phase in a spin-2 BEC.
Spontaneous magnetization arises in the states except for A, B, D, E, and Q (see Table.~\ref{table:spin3-2}).

\begin{table*}
\begin{tabular}{lllll}
 $H'$\hspace{7mm}     &phase\hspace{3mm} & isotropy group $H, \tilde{H}$ & order parameter $\bm\zeta^T$                                      & $\tilde{\mathcal{E}}^{(3)}$ \\[2mm] \hline\hline
 $SO(2)$   &FF    & Eq.~\eqref{eq:Hspin3F3} & $(1,0,0,0,0,0,0)$                                                  & $9c_\gamma$ \\[2mm]
 $SO(2)$   &F     & Eq.~\eqref{eq:Hspin3F2} & $(0,1,0,0,0,0,0)$                                                  & $4c_\gamma$ \\[2mm]
 $SO(2)$   &P     & Eq.~\eqref{eq:Hspin3F1} & $(0,0,1,0,0,0,0)$                                                  & $c_\gamma + 2c_\beta$ \\[2mm]
 $D_\infty$&Q     & Eq.~\eqref{eq:Hspin3P0} & $(0,0,0,1,0,0,0)$                                                  & ${c_\alpha+\frac{4c_\beta}{3}}$ \\[2mm]
 $O$       &D     & $\{e^{i\pi}C_{4z}, e^{i\pi}C_{2,x+z}, e^{i\pi}\mathcal{T}\}$ & $(0,1,0,0,0,-1,0)/\sqrt{2}$                                       & $c_\alpha$ \\[2mm]
 $D_6$     &A     & $\{e^{i\pi}C_{6z},e^{i\pi}C_{2x},e^{i\pi}\mathcal{T}\}$ & $(1,0,0,0,0,0,1)/\sqrt{2}$                                       & ${c_\alpha+\frac{25c_\beta}{12}}$ \\[2mm]
 $C_5$     &H     & $\{e^{-i4\pi/5}C_{5z},e^{i3\pi/5}U_2^{\pi/10}\mathcal{T}\}$ & ${(\sqrt{\frac{2+\eta}{5}},0,0,0,0,\sqrt{\frac{3-\eta}{5}},0)}$  & ${\frac{c_\beta(72c_\gamma-25c_\beta)}{12(3c_\gamma-c_\beta)}}$  \\[2mm] 
           &      &   & $\eta={\frac{c_\beta}{2(c_\beta-3c_\gamma)}}$                 &  \\[2mm] 
 $C_4$     &J     & $\{e^{i\pi/2}C_{4z},C_{2x}\mathcal{T}\}$ & ${(\sqrt{\frac{1+\eta}{4}},0,0,0,\sqrt{\frac{3-\eta}{4}},0,0)}$  & ${\frac{c_\beta(252c_\gamma-25c_\beta)}{12(12c_\gamma-c_\beta)}}$  \\[2mm]
           &      &  & $\eta={\frac{2c_\beta}{12c_\gamma-c_\beta}}$  \\[2mm]
 $D_3$     &E     & $\{C_{3z},e^{i\pi}C_{2x},e^{i\pi}U_2^{\pi/6}\mathcal{T}\}$ & ${(\sqrt{\frac{1-\eta}{4}},0,0,\sqrt{\frac{1+\eta}{2}},0,0,\sqrt{\frac{1-\eta}{4}})}$ & ${\frac{81c_\alpha c_\beta}{48c_\alpha+c_\beta}}$\\[2mm]
           &      & & $\eta={\frac{9c_\beta}{48c_\alpha+c_\beta}}$ \\[2mm]
 $C_3$     &I     & $\{e^{-i2\pi/3}C_{3z},e^{-i\pi/3}U_2^{\pi/6}\mathcal{T}\}$ & ${(0,\sqrt{\frac{1+\eta}{3}},0,0,\sqrt{\frac{2-\eta}{3}},0,0)}$ & ${\frac{c_\beta(72c_\gamma-25c_\beta)}{12(3c_\gamma-c_\beta)}}$\\[2mm]
           &      &  & $\eta={\frac{c_\beta}{2(3c_\gamma-c_\beta)}}$  \\[2mm]
 $C_3$     &R     & $\{C_{3z},e^{i\pi}U_2^{\pi/6}\mathcal{T}\}$ & $(a_+,0,0,\sqrt{\eta},0,0,a_-)$ & ${\frac{c_\gamma(12c_\gamma c_\alpha + 16c_\gamma c_\beta - 9 c_\alpha c_\beta)}{12 c_\gamma^2 - c_\gamma c_\beta - c_\alpha c_\beta}}$\\[2mm]
           &      & & $a_\pm=\sqrt{\frac{1-\eta-\xi}{2}}\pm\sqrt{\frac{\xi}{2}}$\\[2mm]
           &      & & $\eta={\frac{c_\gamma(108c_\gamma - 12c_\alpha - 25c_\beta)}{9(12 c_\gamma^2 - c_\gamma c_\beta - c_\alpha c_\beta)}}$ \\[2mm]
           &      & & $\xi={\frac{24c_\gamma c_\alpha -4c_\gamma c_\beta - 9 c_\alpha c_\beta}{18(12 c_\gamma^2 - c_\gamma c_\beta - c_\alpha c_\beta)}}$ \\ [2mm]
 $D_2$     &B     & $\{C_{2z},e^{i\pi}C_{2x},e^{i\pi}C_{4z}\mathcal{T}\}$ & ${(0,\sqrt{\frac{1-\eta}{4}},0,i\sqrt{\frac{1+\eta}{2}},0,\sqrt{\frac{1-\eta}{4}},0)}$ & ${\frac{c_\beta (18c_\alpha -25 c_\beta)}{3(3c_\alpha-4c_\beta)}}$\\[2mm]
           &      & & $\eta={\frac{c_\beta}{4c_\beta-3c_\alpha}}$ \\[2mm]
 $C_2$     &G     & $\{C_{2z},e^{i\pi}C_{2x}\mathcal{T}\}$ & $(0,a_+,0,\sqrt{\eta},0,a_-,0)$ & ${\frac{25 c_\alpha c_\beta^2 - 4 c_\beta c_\gamma(3 c_\alpha + 25 c_\beta)  + 12 (3 c_\alpha + 4 c_\beta) c_\gamma^2}{12 (c_\alpha c_\beta -4 c_\gamma c_\beta + 3 c_\gamma^2)}}$ \\ [2mm]
           &      & & $a_\pm=\sqrt{\frac{1-\eta-\xi}{2}}\pm\sqrt{\frac{\xi}{2}}$\\[2mm]
           &      & & $\eta={\frac{(5 c_\beta - 6 c_\gamma) (c_\alpha - 4 c_\gamma)} {8 (c_\alpha c_\beta -4c_\beta  c_\gamma + 3 c_\gamma^2)}}$ \\[2mm]
           &      & & $\xi={\frac{c_\beta (2c_\alpha - 3 c_\gamma)}{4 (c_\alpha c_\beta -4c_\beta  c_\gamma + 3 c_\gamma^2)} }$ \\[2mm]
 $C_2$     &C     & $\{e^{i\pi}C_{2z},C_{2x}\mathcal{T}\}$ & $(a,0,b,0,c,0,d)$ & numerically calculated \\[2mm]
           &      & & $a,b,c,d\in \mathbb{R},\ a,b,d>0, c<0$ \\ \hline
\end{tabular}
\caption{Stationary states and mean-field energies for spin-3 BECs obtained by the symmetry-classification method.
The second column indicates the phases discussed in Refs.~\cite{Diener2006,Barnett2007} except for phases J, P, Q, and R.
State I in Ref.~\cite{Diener2006} is refereed to as HH in Ref.~\cite{Barnett2007}. 
P, Q, and R do not appear in the phase diagram of Fig.~\ref{fig:spin3PD}.
The third column shows the isotropy groups.
In A, D, and Q, the time-reversal operator $\mathcal{T}$ is decoupled from spin rotations in $\tilde{H}$,
which implies that only these states have the time-reversal symmetry. 
In the calculation for $D_3$ and $D_2$, we have obtained the following states:
$(\sqrt{2},0,0,\pm i\sqrt{5},0,0,\sqrt{2})^T/3$,
$ (0, \sqrt{3}, 0, \pm \sqrt{10}, 0, \sqrt{3}, 0)^T/4$,
$ (\sqrt{5},0,\sqrt{3},0,\sqrt{3},0,\sqrt{5})^T/4$,
$ (\sqrt{3},0,-\sqrt{5},0,-\sqrt{5},0,\sqrt{3})^T/4$, and
$ (1,0,\sqrt{15},0,\sqrt{15},0,1)^T/4\sqrt{2}$,
which have symmetries of $O$, $D_6$, $D_\infty$, $O$, and $D_6$, respectively.
For the $D_2$ symmetry, we have also obtained the solution in the form of $(a,0,be^{i\delta},0,be^{i\delta},0,a)^T$ ($a,b,\delta\in \mathbb{R}$).
However, this order parameter coincides with that of B by a $\pi/2$ rotation about the $x$ axis.
}
\label{table:spin3}
\end{table*}

\subsubsection{Phase Diagram}
\begin{figure}[ht]
\includegraphics[scale=0.4]{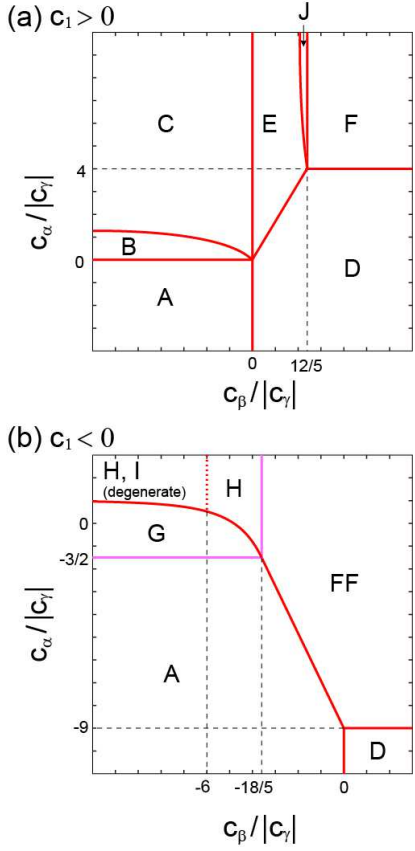}
\caption{(Color online) Phase diagrams of spin-3 BECs for (a) $c_\gamma>0$ and (b) $c_\gamma<0$.
The light-colored lines in (b) indicate second-order phase boundaries, while others show first-order phase boundaries.
In (b), state H is degenerate with state I which can exist only when $c_\beta/|c_\gamma|<-6$.
}
\label{fig:spin3PD}
\end{figure}

By comparing the energies, we obtain the phase diagram of spin-3 spinor BECs as shown in Fig.~\ref{fig:spin3PD}.
Here, the energy of state C is calculated numerically.
The phase diagram is almost consistent with that in Ref.~\cite{Diener2006}.
However, we have found a new phase J with the $C_4$ symmetry, which has eluded the previous works~\cite{Diener2006,Barnett2007}.
We have also investigated the phase diagram by directly solving Eq.~\eqref{eq:stationary}, and confirmed that
no additional phase arises in the phase diagram.
Because the method presented in the present paper can deal with only the states that possess remaining symmetries,
all the phases shown in Fig.~\ref{fig:spin3PD} have certain remaining symmetries.

Figure~\ref{fig:spin3PD}(a) shows the phase diagram for $c_\gamma>0$.
The phase boundary between E and D is given by $y=5x/3$,
while that between phase E and J is $y=(252x-25x^2)/[12(19x-36)]$,
where $y\equiv c_\alpha/|c_\gamma|$ and $x\equiv c_\beta/|c_\gamma|$.
The B--C phase boundary is numerically obtained and well described by
$y=3|x|/(2|x|+3)$.
(The phase boundary $y=3|x|/(3|x|+2)$ given in Ref.~\cite{Diener2006} should read $y=3|x|/(2|x|+3)$.
Figure 1 in Ref.~\cite{Diener2006} agrees with the latter one.)

In the phase diagram for $c_\gamma<0$ [Fig.~\ref{fig:spin3PD}(b)],
the phase boundary between H and G is $y=(24x+5x^2)/(36+24x+5x^2)$, and that between A and FF is $y=-25x/12-9$.
These results also agree with Ref.~\cite{Diener2006}.
In the top left region of Fig.~\ref{fig:spin3PD}(b), states H and I, which have different symmetries, are degenerate.
Distinct from the nematic phase in a spin-2 BEC [Eq.~\eqref{eq:spin2D2}], there is no intermediate state between the H and I states,
as pointed out in Ref.~\cite{Barnett2007}.

\begin{figure}[ht]
\includegraphics[scale=0.4]{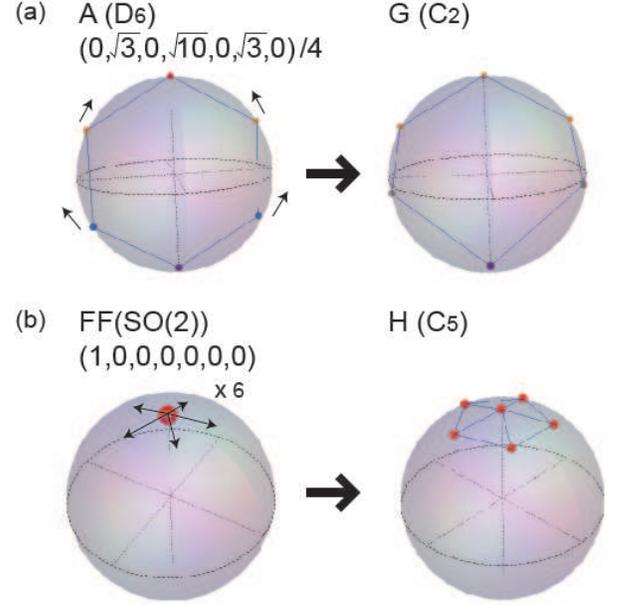}
\caption{(Color online) Symmetry change at (a) A--G and (b) FF--H phase boundaries.
In phase G, the four vertices move upward as indicated by the arrows and the $D_6$ symmetry in phase A breaks down.
However, the six vertices are on the same plane and the $C_2$ symmetry remains in phase G.
In phase H, five vertices, which are on the north pole in phase FF, move downward while keeping the $C_5$ symmetry.
}
\label{fig:spin3-2}
\end{figure}

The phase boundaries between G and A and between FF and H are second-order because they can be transformed into each other by continuous changes in symmetry.
Figure~\ref{fig:spin3-2} shows the continuous symmetry change from A to G, and FF to H.
In phase A, the order parameter has the $D_6$ symmetry.
On the A--G phase boundary, the order parameter of G becomes $(0,\sqrt{3},0,\sqrt{10},0,\sqrt{3},0)^T/4$, where $\eta=5/8$, and $\xi=0$. This state has the $D_6$ symmetry as shown in Fig.~\ref{fig:spin3-2}(a). As $c_\alpha/|c_\gamma|$ increases, 
the four vertices move upward as indicated by arrows in Fig.~\ref{fig:spin3-2}(a), and the $D_6$ symmetry breaks down.
However, the six vertices of G are still on the same plane and the $C_2$ symmetry remains.
On the other hand, on the FF--H phase boundary, the order parameter of H becomes $(1,0,0,0,0,0,0)^T$ which is identical to FF.
In the Majorana representation, the six vertices of FF lie at the north pole.
As $c_\beta/|c_\gamma|$ decreases, one of the vertices remains at the north pole, and the other five move downwards while keeping the $C_5$ symmetry [Fig.~\ref{fig:spin3-2}(b)].

\begin{figure}[htb]
\includegraphics[scale=0.35]{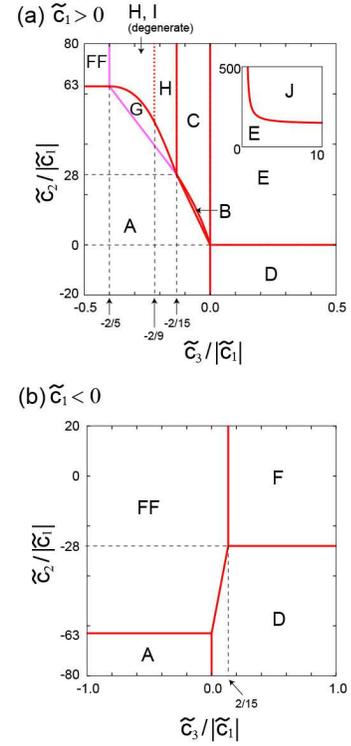}
\caption{(Color online) Phase diagram in the parameter space of 
$(\tilde{c}_3/|\tilde{c}_1|,\tilde{c}_2/|\tilde{c}_1|)$ for (a) $\tilde{c}_1>0$ and (b) $\tilde{c}_1<0$,
obtained by transforming the parameters in Fig.~\ref{fig:spin3PD} according to Eq.~\eqref{eq:cc3}.
In (a), the J phase exists in the region of large $c_2/|c_1|$ and large $c_3/|c_1|$, as shown in the inset of (a).
In the region on the right of FF in (a), H is degenerate with I which can exist only for $\tilde{c}_3/|\tilde{c}_1|<-2/9$.
The curved phase boundaries in (a) are given by B--C: $\tilde{y}=168\tilde{x}(4-15\tilde{x})/(21\tilde{x}-2)$, G--H: $\tilde{y}=252\tilde{x}(5\tilde{x}-2)/(45\tilde{x}^2+12\tilde{x}+4)$,
and E--J: $\tilde{y}=567\tilde{x}/(4\tilde{x}-2)$,
where $\tilde{x}=\tilde{c}_3/|\tilde{c}_1|$ and $\tilde{y}=\tilde{c}_2/|\tilde{c}_1|$.
The light-colored lines in (a) indicate second-order phase boundaries, while others show first-order phase boundaries.
Phases A, D, and FF occupy the bottom left, bottom right, and top left regions, respectively, for both $\tilde{c}_1>0$ and $\tilde{c}_1<0$, because $|\bm f|$, $|A_{00}|$, 
and ${\rm Tr}\,\mathcal{N}^2$ take their minima or maxima in these phases (see Table~\ref{table:spin3-2}).
}
\label{fig:spin3PD2}
\end{figure}

To discuss the underlying physics of the phase digram, we use the mean-field energy in the form of Eq.~\eqref{eq:E3'}.
[We used Eq.~\eqref{eq:E3} for the calculation,
because the description in terms of the order parameter is simpler for Eq.~\eqref{eq:E3} than that for Eq.~\eqref{eq:E3'}.]
In Fig.~\ref{fig:spin3PD2}, we show the phase diagram in the parameter space of $(\tilde{c}_3/|\tilde{c}_1|,\tilde{c}_2/|\tilde{c}_1|)$ for (a) $\tilde{c}_1>0$ and (b) $\tilde{c}_1<0$.
For the case of $F=3$, 
the phase diagram is determined by three terms: $\tilde{c}_1|\bm f|^2$, $\tilde{c}_2|A_{00}|^2$, and $\tilde{c}_3{\rm Tr}\,\mathcal{N}^2$,
whose values for the obtained stationary states are summarized in Table~\ref{table:spin3-2}.
As we discussed in Sec.~\ref{sec:spin2PD}, $|A_{00}|$ takes its maximum when the order parameter has the time-reversal symmetry,
that is, $|A_{00}|=1/\sqrt{7}$ for A, D and Q.
These states may become a ground state for $\tilde{c}_2\to -\infty$.
They differ in the values of ${\rm Tr}\,\mathcal{N}^2$.
Since $\mathcal{N}$ is a $3\times 3$ real symmetric matrix with trace $F(F+1)=12$, it has three real eigenvalues, $\lambda_i\ (i=1,2,3)$, which satisfy $\sum_{i=1,2,3}\lambda_i=12$.
Then, ${\rm Tr}\,\mathcal{N}^2=\sum_{i=1,2,3}\lambda_i^2$ takes its minimum when all three eigenvalues are the same, i.e., ${\rm Tr}\,\mathcal{N}^2|_{\rm min}=48$ for $(\lambda_1,\lambda_2,\lambda_3)=(4,4,4)$.
This is the case of the D phase.
On the other hand, by noting that $0\le \lambda_i\le F^2=9$,
the maximum value ${\rm Tr}\,\mathcal{N}^2|_{\rm max}=171/2$ is achieved in the A phase in which the eigenvalues are given by $(\lambda_1,\lambda_2,\lambda_3)=(3/2,3/2,9)$. 
Hence, phases A and D arise for $\tilde{c}_3<0$ and $\tilde{c}_3>0$, respectively,
whereas phase Q does not appear in the phase diagram because its energy is always between those of A and D.
Note that $\mathcal{N}_{\mu\nu}-f_\mu f_\nu$ is the spin fluctuation, and hence,
${\rm Tr}\,\mathcal{N}^2$ reflects the anisotropy of the spin fluctuation.
The spin fluctuation is isotropic in phase A, while it is most anisotropic in phase D.

In the region of $\tilde{c}_2/|\tilde{c}_1|\to +\infty$ and $\tilde{c}_3/|\tilde{c}_1|\to -\infty$,
the phase that has the minimum $|A_{00}|$ and the maximum ${\rm Tr}\,\mathcal{N}^2$ becomes the ground state, i.e., the FF phase.
Interestingly, ${\rm Tr}\,\mathcal{N}^2$ becomes minimal for the F phase, although the magnetization is the second largest in this phase.
Therefore, for the case of $\tilde{c}_1<0$, the F phase arises in the region of $\tilde{c}_2/|\tilde{c}_1|\to +\infty$ and $\tilde{c}_3/|\tilde{c}_1|\to +\infty$.
On the other hand, there is no state that minimizes all $|\bm f|$, $|A_{00}|$, and ${\rm Tr}\,\mathcal{N}^2$, simultaneously.
Hence, many phases arise in the top right region of Fig.~\ref{fig:spin3PD2}(a).
In the limit of $\tilde{c}_2/|\tilde{c}_1|\to +\infty$ and $\tilde{c}_3/|\tilde{c}_1|\to +\infty$,
the J phase, which has $|\bm f|\to 1/2$, $|A_{00}|=0$, and ${\rm Tr}\,\mathcal{N}^2\to 48$, becomes the ground state.

\begin{table}
\begin{tabular}{llll}\hline
 phase & $|\bm f|$ & $\sqrt{7}|A_{00}|$ & ${\rm Tr}\,\mathcal{N}^2$ \\ \hline\hline
 A     & 0         & $1$ & 171/2\\ 
 B     & 0         & $|\eta|$ & $6(9+2\eta+\eta^2)$\\ 
 C     & \multicolumn{3}{l}{numerically calculated} \\ 
 D     & 0         & $1$ & 48\\ 
 E     & 0         & $|\eta|$ & $9(43-6\eta+27\eta^2)/8$ \\ 
 FF    & 3         & 0 & 171/2\\ 
 F     & 2         & 0 & 48\\ 
 G     & $4\sqrt{\xi(1-\eta-\xi)}$ & $|1-2\xi|$ & $24[2-4\eta^2+5\eta(1-\xi)]$\\ 
 H     & $\eta$    & 0 & $3(36+4\eta+\eta^2)/2$\\ 
 I(HH) & $\eta$    & 0 & $3(36-4\eta+\eta^2)/2$ \\ 
 J     & $\eta$    & 0 & $(99-12\eta+12\eta^2)/2$\\ 
 P     & 1         & 0 & $123/2$\\ 
 Q     & 0         & 1 & $72$\\ 
 R     & $6\sqrt{\xi(1-\eta-\xi)}$ & $|1-2\eta-2\xi|$ & $9(19-30\eta+27\eta^2)/2$\\  \hline
\end{tabular}
\caption{Magnetization $|\bm f|$, singlet-pair amplitude $|A_{00}|$, and nematic tensor ${\rm Tr}\,\mathcal{N}^2$ for the obtained stationary states.
}
\label{table:spin3-2}
\end{table}

\subsubsection{Vortices}
Vortices in spin-3 BECs have been classified in Refs.~\cite{Barnett2007, Yip2007}.
As explained in Refs.~\cite{Barnett2007,Yip2007}, 
topologically stable vortices are classified in terms of 
the elements of the ``lifted'' isotropy group,
which is a subgroup of the universal covering space of $G$, i.e., $SU(2)\times \mathbb{R} \times \Theta$.
Using the symmetry-classification method, we can identify the isotropy group of the obtained state,
and therefore, we can find what types of topological excitations can be hosted there.

The order parameter far from a vortex core is generally described using gauge-transformation and spin-rotation operators as
\begin{align}
\bm\zeta(s)=e^{i\phi(s)}e^{-iF_z\alpha(s)}e^{-iF_y\beta(s)}e^{-iF_z\gamma(s)}\bm\zeta_{\rm rep},
\end{align}
where $0\le s\le 1$ is a parameter describing a closed contour around a vortex,
and $\bm\zeta_{\rm rep}$ is a characteristic order parameter for a state under consideration
(see, for example, the order parameters in the fourth column of Table~\ref{table:spin3}).
For simplicity, we choose $\phi(0)=\alpha(0)=\beta(0)=\gamma(0)=0$.
Then, from the single-valuedness condition for the order parameter [$\bm\zeta(0)=\bm\zeta(1)$],
the operator $e^{i\phi(1)}e^{-iF_z\alpha(1)}e^{-iF_y\beta(1)}e^{-iF_z\gamma(1)}$ must be an element of the isotropy group $H$.

It is worth investigating the mass circulation of a condensate, which is always quantized in a scalar BEC in units of $h/M$.
In spinor BECs, mass circulation is not always quantized due to the contribution from the Berry phase caused by spin textures.
The mass current of a spinor BEC is defined as
\begin{align}
\bm v_{\rm mass} \equiv \frac{\hbar}{2Mi}\sum_m\left[\zeta_m^*(\nabla\zeta_m)-(\nabla\zeta_m^*)\zeta_m\right],
\end{align}
and its circulation is defined as a line integral of $\bm v_{\rm mass}$ along a closed contour $C$:
\begin{align}
\kappa\equiv \oint_C \bm v_{\rm mass} \cdot d\bm \ell.
\end{align}

In phases A, B, D, E, and Q, BECs have no magnetization, and mass current is proportional to the gradient of the overall phase $\phi$:
\begin{align}
\bm v_{\rm mass}=\frac{\hbar}{M}\nabla\phi.
\end{align}
Moreover, these phases have the spin-gauge coupled $\mathbb{Z}_2$ symmetry, namely, the order parameter is invariant under $e^{i\pi} C_{2x}$ or $e^{i\pi}C_{2,x+z}$.
Then, the single-valuedness condition for these states requires that 
$\phi(1)$ takes on an integer multiple of $\pi$.
It follows that $\kappa$ is quantized in units of $h/(2M)$, which is one half of the conventional value.
This situation is similar to the case of a half-quantum vortex in the spin-1 polar BEC~\cite{Zhou2001}. 
Note, however, several topologically different vortices may have the same circulation.
In particular, the A, B, D, and E phases host non-Abelian vortices because their isotropy groups are non-Abelian.
Therefore, a topological charge that classifies each vortex in these phases is an operator (or a matrix) rather than a scalar quantity,
as in the case of a 1/3 vortex in the spin-2 cyclic BEC~\cite{Makela2003,Semenoff2007,Kobayashi2009}.

On the other hand, the other phases in Table~\ref{table:spin3-2} have spontaneous magnetizations, and the mass circulation is not simply quantized.
For these phases, 
the mass current is calculated as
\begin{align}
\bm v_{\rm mass}
&=\frac{\hbar}{M}\left[\nabla\phi - |\bm f|(\cos\beta\nabla\alpha+\nabla\gamma) \right],
\label{eq:vmass_F}
\end{align}
where $|\bm f|$ is the amplitude of the spontaneous magnetization given in the second column of Table.~\ref{table:spin3-2}, and $(\sin\beta\cos\alpha, \sin\beta\sin\alpha, \cos\beta)$ describes the direction of $\bm f$.
Integrating Eq.~\eqref{eq:vmass_F} along a closed contour $C$, we obtain the following relation:
\begin{align}
\frac{M}{\hbar}\kappa -S(\bm f)= \phi(1) - |\bm f|[\alpha(1)+\gamma(1)] \equiv 2\pi I_{\rm v},
\label{eq:spin3_circulation}
\end{align}
where
\begin{align}
 S(\bm f) \equiv |\bm f|\oint_C (1-\cos\beta)\nabla\alpha\cdot d\bm \ell
\end{align}
is the Berry phase due to a texture of $\bm f$,
and it is defined modulo $4\pi|\bm f|$.
Since $\alpha$ and $\beta$ specify the direction of the magnetization, they have to satisfy $\alpha(1)=2\pi n_\alpha$ and $\beta(1)=0$, where $n_\alpha$ is an integer.
For the case of FF, F, and P, there is no discrete symmetry and the single-valuedness condition dictates that
$\phi(1)=2\pi n_\phi$ and $\gamma(1)=2\pi n_\gamma$, where $n_{\phi}$ and $n_\gamma$ are integers. 
Since $|\bm f|$ is an integer in these phases, $I_{\rm v}$ defined in Eq.~\eqref{eq:spin3_circulation} should also be an integer.
Due to the arbitrariness of the Berry phase, vortices are classified by integers mod $2|\bm f|$.
For G, which has spin-gauge decoupled discrete symmetry, 
the single-valuedness condition leads to
$\phi(1)=2\pi n_\phi$ and $\gamma(1)=\pi n_\gamma$.
Substituting these values in Eq.~\eqref{eq:spin3_circulation},
we obtain $I_{\rm v} = n_1 - n_2 |\bm f|/2$ where $n_1$ and $n_2$ are integers ($n_1=n_\phi$ and $n_2=2n_\alpha+n_\gamma$ for this case).
Taking into account the arbitrariness of the Berry phase, vortices in G are classified by a set of two indices, $n_1$ and $n_2$ mod 4.
In a similar manner, we obtain $I_{\rm v}=n_1 - n_2|\bm f| /3$ for the R phase. Topologically distinct vortices are classified by $n_1$ and $n_2$ mod 6. 
For C, H, I, and J, $I_{\rm v}$ is described with three integers due to spin-gauge coupled discrete symmetries.
For example, for the case of H, the single-valuedness condition requires
$\phi(1)=2\pi n_\phi - 4\pi n_3/5$ and $\gamma(1)=2\pi n_\gamma + 2\pi n_3/5$,
leading to $I_{\rm v}=n_1- n_2|\bm f| - n_3(2+|\bm f|)/5$.
A vortex in H is then characterized with a set of three indices, $n_1$, $n_2$ mod 2, and $n_3$ mod 5.
Note that the value of $|\bm f|$ in the above cases (phases G, R, C, H, I, and J) varies, depending on the interaction parameters via parameters $\eta$ and $\xi$ (see Tables~\ref{table:spin3} and \ref{table:spin3-2}).
Hence, the quantization unit of $M\kappa/\hbar -S({\bm f})$ depends on the interatomic interaction.

\section{Conclusion}
\label{sec:conclusions}

We have discussed the symmetry-classification method based on Michel's theorem,
and applied it to spin-1, 2, and 3 spinor Bose-Einstein condensates (BECs).
We classify BECs having unbroken symmetries according to conjugacy classes of an isotropy group $H$,
where $H$ is a group of operations that leave the order parameter unchanged.
For the case of spinor BECs, $H$ is a subgroup of $G=U(1)\times SO(3) \times \Theta$, where $U(1)$, $SO(3)$, and $\Theta$
denote gauge transformations, spin rotations, and time reversal.
For each subgroup $H$ of $G$, we find an order parameter which is invariant under all elements of $H$.
When $H$ is large enough, the order parameter is uniquely determined (inert state). The obtained state is stationary regardless of the detailed form of the interaction energy. 
On the other hand, when $H$ is a small group, there exist many states that are invariant under $H$.
We have characterized the order parameters of these states with a few parameters and found stationary points of the mean-field energy with respect to these parameters.
The obtained order parameter depends on the interaction parameters in the mean-field energy (non-inert state).

For spin-1 and 2 BECs, all ground-state phases are inert states,
except for the spin-2 nematic (antiferromagnetic) phase in which two inert states and the intermediate state between them are degenerate.
For spin-3 BECs, there are fourteen stationary states.
Among them, eleven states appear in the phase diagram: four of them are inert states and others are non-inert states.
We have analytically obtained the order parameters of all stationary states, except for the C phase, as functions of the interaction parameters.
By comparing the energies of the obtained states, 
we have found a new phase (J phase) which exists in a very narrow region in the parameter space and has eluded the previous works~\cite{Diener2006,Barnett2007}.

Using the symmetry-classification method, we can find the isotropy group of the obtained state,
from which we see which types of topological excitations can be hosted in the phase.
Among the obtained stationary states, A, B, D, and E host non-Abelian vortices since their isotropy groups are non-Abelian.
The mass circulation of these states and that of Q are quantized in units of $h/(2M)$.
In the other phases,
the BEC has nonzero magnetization and the circulation of mass current is not quantized, due to the contribution from the Berry phase caused by spin textures.
The difference between mass circulation and the Berry phase, however, is quantized as in Eq.~\eqref{eq:spin3_circulation}.

It is impossible to find a ground state that has no remaining symmetry, i.e., $H=1$, by using the symmetry-classification method.
For the case of $H=1$, the symmetry-classification method amounts to solving Eq.~\eqref{eq:stationary}.
For the case of spin-1, 2 and 3 BECs in the absence of an external field, all ground states have remaining symmetries and are found by the symmetry-classification method.
However, in the presence of an external field, 
the full symmetry $G$ becomes smaller and completely broken in some phases, such as the broken-axisymmetry phase in a spin-1 BEC and the $Z_1, Z_2$ and $Z_3$ phases in spin-3 BECs~\cite{Diener2006}.

\acknowledgments 
This work was supported by MEXT
(KAKENHI 22740265 and 22340114, 
a Grant-in-Aid for Scientific Research on Innovation Areas ``Topological Quantum Phenomena" (KAKENHI 22103005),
a Global COE Program ``the Physical Sciences Frontier,"
and the Photon Frontier Network Program),
and JSPS and FRST under the Japan-New Zealand Research Cooperative Program.

\appendix
\section{Stationary states with discrete symmetries in spin-3 BECs}
\label{sec:append_spin3}

In this appendix, we explore stationary states
which have $D_n$ and $C_n$ symmetries in spin-3 BECs.

\subsection{$D_6$ symmetry}
The matrix representation of $C_{6z}$ and $C_{2x}$ are given by
\begin{align}
C_{6z}&={\rm Diag}[-1,e^{-i2\pi/3},e^{-i\pi/3},1,\nonumber\\
&\hspace{30mm}e^{i\pi/3},e^{i2\pi/3},-1],\\
C_{2x}&=\begin{pmatrix}
0 & 0 & 0 & 0 & 0 & 0 &-1 \\
0 & 0 & 0 & 0 & 0 &-1 & 0 \\
0 & 0 & 0 & 0 &-1 & 0 & 0 \\
0 & 0 & 0 &-1 & 0 & 0 & 0 \\
0 & 0 &-1 & 0 & 0 & 0 & 0 \\
0 &-1 & 0 & 0 & 0 & 0 & 0 \\
-1 & 0 & 0 & 0 & 0 & 0 & 0 \\
\end{pmatrix}.
\end{align}
The simultaneous eigenstate of these operators is uniquely determined to be
\begin{align}
(\frac{1}{\sqrt{2}},0,0,0,0,0,\frac{1}{\sqrt{2}})^T,
\end{align}
where eigenvalues of $C_{6z}$ and $C_{2x}$ are both equal to $-1$.
The generators of the isotropy group and the mean-field energy for this state are given by
\begin{align}
&\tilde{H}=\{e^{i\pi} C_{6z}, e^{i\pi} C_{2x}, e^{i\pi}\mathcal{T} \},\\
&\tilde{\mathcal{E}}^{(3)}= c_\alpha + \frac{25}{12}c_\beta,
\end{align}
respectively.

\subsection{$C_6$ symmetry}
A nontrivial eigenstate of $C_{6z}$, namely, the eigenstate that has more than two nonzero components of the order parameter,
is written as
\begin{align}
(a,0,0,0,0,0,b)^T,
\label{eq:spin3C6}
\end{align}
where $a$ and $b$ are complex numbers with $|a|^2+|b|^2=1$.
Note that we can arbitrarily choose the phases of $a$ and $b$
by applying a gauge transformation and a spin rotation about the $z$ axis.
Here, we choose $a$ and $b$ to be real positive numbers and write $a=\sqrt{1-\eta}$ and $b=\sqrt{\eta}$, where $0<\eta\le 1/2$.
Then, the mean-field energy of this state can be written as a function of $\eta$ as
\begin{align}
\tilde{\mathcal{E}}^{(3)}=
9 c_\gamma (1 - 2 \eta)^2 + \frac{12 c_\alpha + 25 c_\beta}{3}\eta (1-\eta).
\label{eq:ene_spin3C6}
\end{align}
The stationary point of Eq.~\eqref{eq:ene_spin3C6} is $\eta=1/2$.
Hence, the stationary point that has the $C_6$ symmetry always possesses the $D_6$ symmetry.

\subsection{$D_5$ symmetry}
The matrix representation of $C_{5z}$ is given by
\begin{align}
C_{5z}=&{\rm Diag}[e^{-i6\pi/5},e^{-i4\pi/5},e^{-i2\pi/5},\nonumber\\
&\hspace{25mm}1,e^{i2\pi/5},e^{i4\pi/5},e^{i6\pi/5}].
\end{align}
There is no simultaneous eigenstate of $C_{5z}$ and $C_{2x}$.

\subsection{$C_5$ symmetry}
In a manner similar to the case of the $C_6$ symmetry,
a nontrivial eigenstate of $C_{5z}$ can be written without loss of generality as
\begin{align}
(\sqrt{\frac{2+\eta}{5}},0,0,0,0,\sqrt{\frac{3-\eta}{5}},0)^T,
\label{eq:spin3C5-0}
\end{align}
where $-2<\eta<3$, and
the eigenvalue is $e^{i4\pi/5}$.
The order parameter in the form of $(0,\sqrt{(3-\eta)/5},0,0,0,0,\sqrt{(2+\eta)/5})^T$ is also a nontrivial eigenstate of $C_{5z}$.
However, this state belongs to the same orbit as that of the state in Eq.~\eqref{eq:spin3C5-0}, because they are transformed into each other by a spin rotation and a gauge transformation.
The mean-field energy is calculated as a function of $\eta$ as
\begin{align}
\tilde{\mathcal{E}}^{(3)}=
 c_\gamma \eta^2 + \frac{c_\beta}{3} (6 + \eta - \eta^2),
\end{align}
which has a stationary point at
\begin{align}
 \eta=\frac{c_\beta}{2(c_\beta-3c_\gamma)}.
\end{align}
The stationary point exists only when $-2<c_\beta/[2(c_\beta-3c_\gamma)]<3$.
The generators of the isotropy group and the mean-field energy for the stationary state are given by
\begin{align}
&\tilde{H}=\{e^{-i4\pi/5} C_{5z}, e^{i3\pi/5}U_2^{\pi/10}\mathcal{T} \},\\
&\tilde{\mathcal{E}}^{(3)}= \frac{c_\beta(72 c_\gamma-25 c_\beta)}{12(3c_\gamma-c_\beta)},
\end{align}
respectively.

\subsection{$D_4$ symmetry}
The matrix representation of $C_{4z}$ is given by Eq.~\eqref{eq:spin3_matrixC4z}.
The simultaneous eigenstate of $C_{4z}$ and $C_{2x}$ is determined up to an overall phase to be
\begin{align}
(0,\frac{1}{\sqrt{2}},0,0,0,\frac{1}{\sqrt{2}},0)^T.
\end{align}
This state has the symmetry of the octahedron,
and the $D_4$ symmetry is not the largest symmetry of this state.

\subsection{$C_4$ symmetry}
There are two nontrivial eigenstates of $C_{4z}$.

{\it Case} (i): The order parameter
\begin{align}
(\sqrt{\frac{1+\eta}{4}},0,0,0,\sqrt{\frac{3-\eta}{4}},0,0)^T,\label{eq:spin3C4-1}
\end{align}
is an eigenstate of $C_{4z}$ with eigenvalue $i$,
where $-1<\eta<3$.
The mean-field energy of this state is written as a function of $\eta$ as
\begin{align}
\tilde{\mathcal{E}}^{(3)}=c_\gamma \eta^2 + \frac{c_\beta}{12} (3 -\eta) (7 +\eta).
\label{eq:ene_spin3C4}
\end{align}
By taking derivative of Eq.~\eqref{eq:ene_spin3C4} with respect to $\eta$, we find a stationary state at
\begin{align}
\eta=\frac{2c_\beta}{12c_\gamma-c_\beta}.
\end{align}
The generators of the isotropy group and the mean-field energy for the stationary state are given by
\begin{align}
&\tilde{H}=\{e^{i\pi/2} C_{4z}, C_{2x}\mathcal{T} \},\\
&\tilde{\mathcal{E}}^{(3)} = \frac{c_\beta(252 c_\gamma-25 c_\beta)}{12(12c_\gamma-c_\beta)},
\end{align}
respectively.

{\it Case} (ii): The order parameter
\begin{align}
(0,\sqrt{\frac{1+\eta}{2}},0,0,0,\sqrt{\frac{1-\eta}{2}},0)^T,\label{eq:spin3C4-2}
\end{align}
is an eigenstate of $C_{4z}$ with eigenvalue $-1$, where $-1<\eta<1$.
The man-field energy of this state is given by
\begin{align}
\tilde{\mathcal{E}}^{(3)}=c_\alpha - 4 (c_\alpha - 4 c_\gamma) \eta^2,
\end{align}
whose stationary point lies at $\eta=0$.
The corresponding state has the symmetry of the octahedron ($O$),
and $C_4$ is not the largest symmetry of this state.

\subsection{$D_3$ symmetry}
The matrix representation of $C_{3z}$ is given by
\begin{align}
C_{3z}={\rm Diag}[1,e^{i2\pi/3},e^{-i2\pi/3},1,e^{i2\pi/3},e^{-i2\pi/3},1].
\end{align}
The simultaneous eigenstate of $C_{3z}$ and $C_{2x}$ is written in the form of
\begin{align}
(\sqrt{\frac{1-\eta}{4}},0,0,e^{i\delta}\sqrt{\frac{1+\eta}{2}},0,0,\sqrt{\frac{1-\eta}{4}})^T,
\end{align}
where $-1<\eta<1$ and the eigenvalues of $C_{3z}$ and $C_{2x}$ are 1 and $-1$, respectively.
The mean-field energy of this state is given by
\begin{align}
\tilde{\mathcal{E}}^{(3)}=& \frac{c_\alpha}{2}(1+\eta^2) + \frac{c_\beta}{48}(41-18\eta+41\eta^2)\nonumber\\
&+ \frac{5c_\beta-3c_\alpha}{6}(1-\eta^2)\cos(2\delta).
\end{align}
The stationary points of this function are obtained as
\begin{align}
{\rm (a)}&\ \eta = \frac{1}{9},\ \delta=\pm \frac{\pi}{2},\\
{\rm (b)}&\ \eta = \frac{9c_\beta}{48c_\alpha+c_\beta},\ \delta=0, \pi.
\end{align}
For case (a), the stationary state is an inert state whose order parameter is given by $(2,0,0,\pm i\sqrt{5},0,0,2)^T/3$.
Investigating the geometric structure of this state, we can find that this state has the symmetry of an octahedron.
On the other hand, for case (b), the the stationary state is a non-inert state.
The order parameters at $\delta=0$ and $\pi$ are
related to each other by a gauge transformation and a spin rotation.
As a result, the stationary state that has the $D_3$ symmetry can be written as
\begin{align}
&(\sqrt{\frac{1-\eta}{4}},0,0,\sqrt{\frac{1+\eta}{2}},0,0,\sqrt{\frac{1-\eta}{4}})^T,\\
&\eta=\frac{9c_\beta}{48c_\alpha+c_\beta}.
\end{align}
The generators of the isotropy group and the mean-field energy for this stationary state are given by
\begin{align}
&\tilde{H}=\{C_{3z}, e^{i\pi}C_{2x}, e^{i\pi}U_2^{\pi/6}\mathcal{T} \},\\
&\tilde{\mathcal{E}}^{(3)} = \frac{81 c_\alpha c_\beta}{48 c_\alpha+c_\beta},
\end{align}
respectively.

\subsection{$C_3$ symmetry}
There are two nontrivial eigenstates of $C_{3z}$.

{\it Case} (i): The order parameter
\begin{align}
(0,\sqrt{\frac{1+\eta}{3}},0,0,\sqrt{\frac{2-\eta}{3}},0,0)^T,\label{eq:spin3C3-1}
\end{align}
is the eigenstate of $C_{3z}$ with eigenvalue $e^{i2\pi/3}$. 
The mean-field energy of this state is written in terms of $\eta$ as
\begin{align}
\tilde{\mathcal{E}}^{(3)}=c_\gamma \eta^2 + \frac{c_\beta}{3}(6-\eta-\eta^2),
\end{align}
which has a stationary point at 
\begin{align}
\eta=\frac{c_\beta}{2(3c_\gamma-c_\beta)}.
\end{align}
The generators of the isotropy group and the mean-field energy for the stationary state are given by
\begin{align}
&\tilde{H}=\{e^{-i2\pi/3}C_{3z},  e^{-i\pi/3}U_2^{\pi/6}\mathcal{T} \},\\
&\tilde{\mathcal{E}}^{(3)} = \frac{c_\beta(72 c_\gamma-25 c_\beta)}{12(3c_\gamma-c_\beta)},
\end{align}
respectively.

{\it Case} (ii): The order parameter
\begin{align}
(a_+,0,0,b,0,0,a_-)^T,\label{eq:spin3C3-2}
\end{align}
is the eigenstate of $C_{3z}$ with eigenvalue $1$,
where $a_\pm$ and $b$ are complex numbers that satisfy $|a_+|^2+|b|^2+|a_-|^2=1$.
Here, we choose $a_\pm$ to be real and rewrite these parameters as
\begin{align}
a_\pm &= \sqrt{\frac{1-\eta-\xi}{2}} \pm \sqrt{\frac{\xi}{2}},\\
b     &= e^{i\delta}\sqrt{\eta},
\end{align}
where $\eta>0$, $\xi\ge 0$, $\eta+\xi<1$, and $-\pi<\delta\le \pi$.
Then, the mean-field energy can be written as a function of $\eta,\xi$ and $\delta$ as
\begin{align}
\tilde{\mathcal{E}}^{(3)} =
&c_\alpha (1-2\eta+2\eta^2-4\xi+4\xi \eta+4\xi^2) \nonumber\\
&+\frac{c_\beta}{12} (25-50\eta+41\eta^2-100\xi+100\xi \eta+100\xi^2)\nonumber\\
&+36 c_\gamma \xi(1-\xi-\eta)\nonumber\\
&-\frac{2(3c_\alpha-5c_\beta)}{3} \eta (1-\eta-2\xi) \cos(2\delta)
\end{align}
which has a stationary point at $\delta=0$ and
\begin{align}
\eta&=\frac{(108 c_\gamma - 12 c_\alpha - 25 c_\beta )c_\gamma}
{9 (12 c_\gamma^2 - c_\alpha c_\beta - c_\beta c_\gamma)},\\
\xi &= \frac{ 24 c_\alpha c_\gamma - 9 c_\alpha c_\beta  - 4 c_\beta c_\gamma}{18(12 c_\gamma^2 - c_\alpha c_\beta  - c_\beta c_\gamma )}.
\end{align}
The generators of the isotropy group and the mean-field energy for the stationary state are given by
\begin{align}
&\tilde{H}=\{C_{3z},  e^{i\pi}U_2^{\pi/6}\mathcal{T} \},\\
&\tilde{\mathcal{E}}^{(3)} = 
\frac{c_\gamma(12c_\gamma c_\alpha+16c_\gamma c_\beta-9c_\alpha c_\beta)}{12c_\gamma^2-c_\gamma c_\beta-c_\alpha c_\beta},
\end{align}
respectively.
We have also obtained stationary points at $\delta=\pm \pi/2$ and $\pi$,
which lie on the same orbit as that of the above stationary state.

\subsection{$D_2$ symmetry}
The matrix representation of $C_{2z}$ is given by
\begin{align}
C_{2z}={\rm Diag}[-1,1,-1,1,-1,1,-1].
\end{align}
There are two simultaneous eigenstates of $C_{2z}$ and $C_{2x}$.

{\it Case} (i):
The order parameter
\begin{align}
(0,\sqrt{\frac{1-\eta}{4}},0,e^{i\delta}\sqrt{\frac{1+\eta}{2}},0,\sqrt{\frac{1-\eta}{4}},0)^T,
\label{eq:spin3D2}
\end{align}
is a simultaneous eigenstate of $C_{2z}$ and $C_{2x}$ with eigenvalues $1$ and $-1$, respectively.
The mean-field energy is given by
\begin{align}
\tilde{\mathcal{E}}^{(3)} = 
&\frac{2c_\beta}{3}(3+\eta-2\eta^2) + \frac{c_\alpha}{2}(1+\eta^2)\nonumber\\
&+\frac{c_\alpha}{2}(1-\eta^2)\cos(2\delta).
\end{align}
By taking derivatives with respect to $\eta$ and $\delta$,
we find stationary states at
\begin{align}
\textrm{(a)}\ &\eta=1/4,\ \delta=0,\pi,\\
\textrm{(b)}\ &\eta=\frac{c_\beta}{4c_\beta-3c_\alpha},\ \delta=\pm \frac{\pi}{2}
\end{align}
For case (a), the stationary state is uniquely determined to be an inert state
whose order parameter is given by $(0,\sqrt{3},0,\pm\sqrt{10},0,\sqrt{3},0)^T/4$.
The Majorana representation of this state has six vertices that form a hexagon with the $D_6$ symmetry.
For case (b), the stationary state is a non-inert state.
The generators of the isotropy group and the mean-field energy for this state are given by
\begin{align}
&\tilde{H}=\{C_{2z},e^{i\pi}C_{2x}, e^{i\pi}C_{4z}\mathcal{T}\}, \label{eq:tildeH_spin3D2}\\
&\tilde{\mathcal{E}}^{(3)} = \frac{c_\beta (18c_\alpha -25 c_\beta)}{3(3c_\alpha-4c_\beta)}.
\end{align}

{\it Case} (ii):
The order parameter
\begin{align}
(\sqrt{\frac{1-\eta}{2}},0,e^{i\delta}\sqrt{\frac{\eta}{2}},0,e^{i\delta}\sqrt{\frac{\eta}{2}},0,\sqrt{\frac{1-\eta}{2}})^T
\end{align}
is a simultaneous eigenstate of $C_{2z}$ and $C_{2x}$ with eigenvalues $-1$ for both operators.
The mean-field energy of this state is given by
\begin{align}
\tilde{\mathcal{E}}^{(3)}=
&c_\alpha(1-2\eta+2\eta^2)+\frac{c_\beta}{12}(25-30\eta+26\eta^2)\nonumber\\
&+8\sqrt{15}c_\beta \sqrt{\eta^3(1-\eta)}\cos\delta\nonumber\\
&+6(4c_\alpha-5c_\beta)\eta(1-\eta)\cos(2\delta),
\end{align}
whose stationary points are obtained as
\begin{align}
{\rm (a)}&\ \eta=\frac{3}{8},\ \delta=0,\\
{\rm (b)}&\ \eta=\frac{5}{8},\ \delta=\pi,\\
{\rm (c)}&\ \eta=\frac{15}{16}\ \delta=0,\\
{\rm (d)}&\ \eta=\frac{3 (4 c_\alpha - 5 c_\beta)}{8 (3 c_\alpha - 4 c_\beta)},\nonumber\\
&\ \delta = \arccos\left[\frac{1-2\eta}{2}\sqrt{\frac{15}{\eta(1-\eta)}}\right].
\end{align}
For cases (a)--(c), the corresponding order parameters are respectively given by
\begin{align}
{\rm (a)}&\ \frac{1}{4}(\sqrt{5},0,\sqrt{3},0,\sqrt{3},0,\sqrt{5})^T,\\
{\rm (b)}&\ \frac{1}{4}(\sqrt{3},0,-\sqrt{5},0,-\sqrt{5},0,\sqrt{3})^T,\\
{\rm (c)}&\ \frac{1}{4\sqrt{2}}(1,0,\sqrt{15},0,\sqrt{15},0,1)^T,
\end{align}
which are all inert states and have the symmetries of $D_\infty$ [$\bm\zeta^{(0)}$ in Eq.~\eqref{eq:spin3P0}] , $O$ and $D_6$, respectively.
For case (d), the generator of this state is given by
\begin{align}
\tilde{H}=\{e^{i\pi}C_{2z}, e^{i\pi}C_{2x}, e^{i\theta}C_{4y}\mathcal{T}\},
\end{align}
where
\begin{align}
 \theta\equiv \pi+\arctan\left(\frac{\sqrt{15\eta}\sin\delta}{\sqrt{1-\eta}+\sqrt{15\eta}\cos\delta}\right).
\end{align}
Since the isotropy group generated by this $\tilde{H}$ is isomorphic to that generated by $\tilde{H}$ in Eq.~\eqref{eq:tildeH_spin3D2}, the order parameter for case (d) lies on the same orbit as that of the state in Eq.~\eqref{eq:spin3D2}.

\subsection{$C_2$ symmetry}
There are two nontrivial eigenstates of $C_2$.

{\it Case} (i): The order parameter
\begin{align}
& (0,a_+,0,e^{i\delta}\sqrt{\eta},0,a_-,0)^T,\\
& a_\pm = \sqrt{\frac{1-\eta-\xi}{2}}\pm\sqrt{\frac{\xi}{2}},
\end{align}
is the eigenstate of $C_2$ with eigenvalue 1,
where $\eta>0$, $\xi\ge 0$, $\eta+\xi<1$, and $-\pi<\delta\le \pi$.
The mean-field energy of this state is given by
\begin{align}
\tilde{\mathcal{E}}^{(3)}=
&\frac{4c_\beta}{3}\eta(5-4\eta)
+c_\alpha(1-2\eta+2\eta^2) \nonumber\\
&-4(c_\alpha-4c_\gamma)\xi(1-\eta-\xi) \nonumber\\
&+2c_\alpha\eta(1-\eta-2\xi)\cos(2\delta),
\end{align}
which has a stationary point at
\begin{align}
\eta &=\frac{(5 c_\beta - 6 c_\gamma) (c_\alpha - 4 c_\gamma)} {8 (c_\alpha c_\beta -4c_\beta  c_\gamma + 3 c_\gamma^2)},\nonumber\\
\xi  &=\frac{c_\beta (2c_\alpha - 3 c_\gamma)}{4 (c_\alpha c_\beta -4c_\beta  c_\gamma + 3 c_\gamma^2)},\nonumber \\
\delta &=0.
\end{align}
The generators of the isotropy group and the mean-field energy for this state are given by
\begin{align}
&\tilde{H}=\{C_{2z},e^{i\pi}C_{2x}\mathcal{T}\},\\
&\tilde{\mathcal{E}}^{(3)}= \frac{25 c_\alpha c_\beta^2 - 4 c_\beta c_\gamma(3 c_\alpha + 25 c_\beta)  + 12 (3 c_\alpha + 4 c_\beta) c_\gamma^2}{12 (c_\alpha c_\beta -4 c_\gamma c_\beta + 3 c_\gamma^2)},
\end{align}
respectively.
We have also obtained the stationary points at $\delta=\pm \pi/2 $ and $\pi$, which lie on the same orbit as that of the above solution.

{\it Case} (ii): The order parameter in the form of
\begin{align}
 (a,0,b,0,c,0,d)^T
\end{align}
is the eigenstate of $C_2$ with eigenvalue $-1$,
where $a, b, c$ and $d$ are complex numbers satisfying $|a|^2+|b|^2+|c|^2+|d|^2=1$.
We have numerically minimized the mean-field energy of this state with respect to $a,b,c$ and $d$.
In the obtained stationary state, we can choose all components in the order parameter to be real and $a, b, d >0$ and $c<0$.


%

\end{document}